\documentclass[twocolumn,preprintnumbers,amsmath,amssymb,prb,superscriptaddress]{revtex4}
\usepackage[xdvi]{graphicx}
\usepackage{latexsym}
\usepackage{amsbsy}
\usepackage{amsmath}
\usepackage{times}
\usepackage{verbatim}
\usepackage{color}
\usepackage{braket}
\usepackage{units}
\usepackage{todonotes}
\usepackage{multirow}

\newcommand{\CrO}{Cr$_2$O$_3$}

\newcommand{\rev}[1]{{\textcolor{black}{#1}}}

\newcommand{\AIR}{A_\mathrm{IR}}
\newcommand{\AR}{A_\mathrm{R}}
\newcommand{\AOM}{A_\mathrm{\Omega}}

\newcommand{\noIR}{\widetilde{\omega}_\mathrm{IR}}
\newcommand{\noR}{\widetilde{\omega}_\mathrm{R}}

\newcommand{\oIR}{\omega_\mathrm{IR}}
\newcommand{\oR}{\omega_\mathrm{R}}

\renewcommand{\vec}[1]{\mbox{\boldmath$#1$}}
\newcommand{\PH}[2]{\ensuremath{#1_{#2}}}

\begin{document}

\date{\today}
\title{Magnetophononics: ultrafast spin control through the lattice}

\author{M. Fechner}
\email{michael.fechner@mpsd.mpg.de}
\affiliation{Max Planck Institute for the Structure and Dynamics of Matter, 22761 Hamburg, Germany}
\affiliation{Materials Theory, ETH Zurich, Wolfgang-Pauli-Strasse 27, 8093 Z\"urich, Switzerland}
\author{A. Sukhov}
\affiliation{Institut f\"ur Physik, Martin-Luther-Universit\"at Halle-Wittenberg, 06099 Halle/Saale, Germany}
\affiliation{Forschungszentrum J\"ulich GmbH, Helmholtz Institute Erlangen-N\"urnberg for Renewable Energy (IEK-11), 90429 N\"urnberg, Germany}
\author{L. Chotorlishvili}
\affiliation{Institut f\"ur Physik, Martin-Luther-Universit\"at Halle-Wittenberg, 06099 Halle/Saale, Germany}
\author{C. Kenel}
\affiliation{Materials Theory, ETH Zurich, Wolfgang-Pauli-Strasse 27, 8093 Z\"urich, Switzerland}
\affiliation{Department of Materials Science and Engineering, McCormick School of Engineering, Northwestern University, 2200 Campus Drive, Evanston, IL 60208, USA}
\author{J. Berakdar}
\affiliation{Institut f\"ur Physik, Martin-Luther-Universit\"at Halle-Wittenberg, 06099 Halle/Saale, Germany}
\author{N. A. Spaldin}
\affiliation{Materials Theory, ETH Zurich, Wolfgang-Pauli-Strasse 27, 8093 Z\"urich, Switzerland}

\begin{abstract}
Using a combination of first-principles and magnetization-dynamics calculations, we study the effect of the intense optical excitation of phonons on the magnetic behavior in insulating magnetic materials. Taking the prototypical magnetoelectric \CrO\ as our model system, we show that excitation of a polar mode at 17 THz causes a pronounced modification of the magnetic exchange interactions through a change in the average Cr-Cr distance. In particular, the quasi-static deformation induced by nonlinear phononic coupling yields a structure with a \rev{modified magnetic} state, which persists for the duration of the phonon excitation. In addition, our time-dependent magnetization dynamics computations show that systematic modulation of the magnetic exchange interaction by the phonon excitation modifies the magnetization dynamics. This temporal modulation of the magnetic \rev{ exchange interaction strengths} using phonons provides a new route to creating non-equilibrium magnetic states and suggests new avenues for fast manipulation of spin arrangements and dynamics. 
\end{abstract}
\maketitle
\section{Introduction}
The field of non-linear phononics, in which high-intensity terahertz (THz) optical pulses are used to drive phonon excitations, is of increasing interest \cite{Forst:2011ep,Subedi:2014ik}. The non-linear processes triggered by the strong phonon excitations have been shown repeatedly to introduce complex structural modifications in materials, which in turn cause striking and often unexpected changes in properties. Examples include the stimulation of insulator to metal transitions in correlated oxides \cite{Rini:2007hca,Caviglia:2012eaa,Esposito:2017wy} and the enhancement of superconducting properties in high-$T_c$ cuprates\cite{Kaiser:2014de,Hu:2014cg} and other materials\cite{Mitrano:2016fr}. In all cases, theoretical studies combining density functional theory with phenomenological modeling have been invaluable in interpreting the experimental results \cite{Subedi:2014ik,Mankowsky:2014vt,Fechner:2016tu,Mankowsky:2017br,Gu:2017dm,Juraschek:2017ic} and even in predicting new phenomena, such as the recent switching\cite{Subedi:2015dw} and creation\cite{Juraschek:2017ur} of ferroic states, ahead of their experimental observation \cite{Mankowsky:2017ur}. 

In addition to modifying electronic properties, there are a number of examples of THz phonon excitation triggering magnetic phenomena on a picosecond (ps$=10^{-12}$s) time-scale. Early results indicate that selective phonon excitations can induce demagnetization processes\cite{Forst:2011kl,Forst:2015fv}, and two-phonon excitation \cite{Nova:2016ja} has been shown to excite magnons by the stimulated rotational motion of atoms\rev{\cite{Juraschek:2017ic,Juraschek:2017ur}}. We note that these behaviors are distinct from ultrafast femtosecond (fs$=10^{-15}$s) spin-flip relaxation processes induced by optical frequency pulses, such as the pioneering experiments of Refs.~[\onlinecite{Beaurepaire:1996es,Kimel:2002ej,Stanciu:2007fy}], which heat the electronic/lattice sub-system. They are also distinct from the THz excitation of electro-magnons in multiferroics\cite{Kubacka:2014bf}, in which the electric field of the light pulse couples directly to the dipole moment of the electron-magnon quasiparticle.

In this work, we address theoretically how the structural changes triggered by the non-linear phononic processes affect the magnetic energy landscape. We are particularly interested in the situation in which an excited infra-red-active phonon mode couples quadratic-linearly to a Raman-active mode, causing a shift in the average structure that persists for the duration of the phonon excitation. We show that the induced structure can have a different magnetic ordering from the equilibrium structure, so that the lattice excitation can cause a spin-state transition. In addition, we explore the spin dynamics induced by the phonon coupling, and show that various complex spin-flip patterns can be selectively excited through appropriate choice of the phonon driving frequency. 

\begin{figure}[b]
\includegraphics[width=86mm]{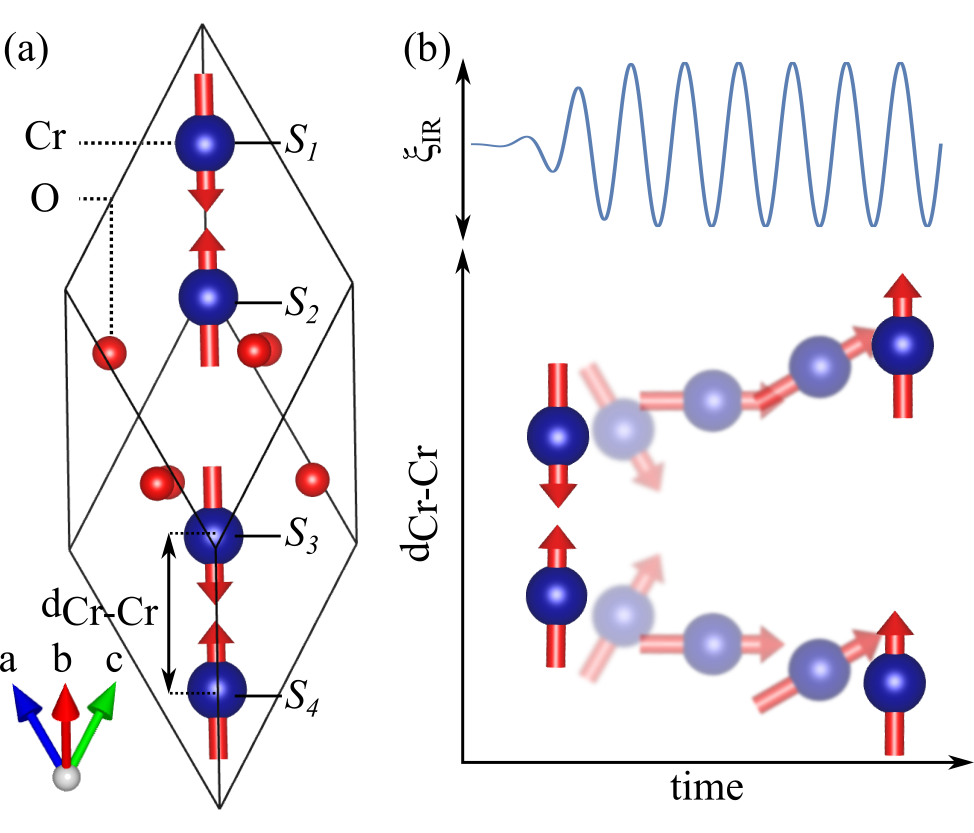}
\caption{\label{fig_1}(a) Unit cell of \CrO\ with the red arrows indicating the ground state antiferromagnetic spin magnetic order. (b) Schematics of the phonon-driven change in magnetic ground state: The excitation of a polar phonon mode ($\xi_\mathrm{IR}$) induces an increase in the nearest-neighbor Cr-Cr bond length by the square-linear anharmonic phonon coupling. The longer bond length results in ferromagnetic exchange interaction between the Cr ions creating a transient \rev{change in the magnetic} state for the duration of the phonon excitation.} 
\end{figure}

In the next section we review the now well-established theory of non-linear phononics. We then present a model that combines the non-linear phononics formalism with the Heisenberg Hamiltonian to describe spin-phonon coupling through the changes in magnetic exchange interactions that are induced by changes in structure. In Section~\ref{Cr2O3}, we apply the model to the prototypical magnetoelectric material, \CrO\  (Fig.~\ref{fig_1}), using first-principles calculations to obtain all the material-specific parameters. In Section~\ref{NLLD} we present and discuss the analytical solution of the non-linear phononic Hamiltonian for the lattice dynamics and in Section~\ref{SpinDynamics} the numerical simulations of the magnetization dynamics based on the Landau-Lifshitz Gilbert equation\cite{Landau:1935tr,Gilbert:1955ta}. The implications of our findings and suggestions for future work are discussed in the Summary.

\section{Theory}

Here we describe separately the modeling of phononic and magnetic lattice systems before outlining our approach to modeling their coupling. We begin with the description of lattice anharmonicity.

For large atomic displacements, such as those induced by intense optical pulses, the usual harmonic description of lattice phonons breaks down and higher order anharmonicities become relevant. The lattice Hamiltonian can then be written as\cite{Subedi:2014ik,Fechner:2016tu} 
\begin{equation}
\begin{aligned}
\label{eq_anharmonic}
H^\mathrm{vib}(\xi_{\mathrm{IR}},\xi_{\mathrm{R}})=&\frac{\omega_{\mathrm{IR}}^2}{2}\xi_{\mathrm{IR}}^2+\frac{\omega_{\mathrm{R}}^2}{2}\xi_{\mathrm{R}}^2 + g\;\xi_\mathrm{R} \xi^2_\mathrm{IR} \\
&+ \frac{\gamma_{\mathrm{IR}}}{4}\xi^4_{\mathrm{IR}}+\frac{\gamma_{\mathrm{R}}}{4}\xi^4_{\mathrm{R}}\quad ,
\end{aligned}
\end{equation}
\rev{where $\omega_{\mathrm{IR}}$, $\omega_{\mathrm{R}}$ are the frequencies of the infrared (IR) and Raman (R) modes, $\xi_{\mathrm{IR}}$ and $\xi_{\mathrm{R}}$ are distortions, \rev{ $\gamma_{\mathrm{IR/R}}$} are fourth order anharmonic constants and $g$ represents the coupling between two phonon modes.} (Terms in $\xi_{\mathrm{R}}^3$ are small and so are neglected for conciseness.) The dominant anharmonic response to optical pumping comes from the third-order $\xi_\mathrm{R} \xi^2_\mathrm{IR}$ term, which has been shown to cause a shift in the potential energy to a finite value of the Raman normal mode coordinate, creating in turn a quasi-static change in the structure \cite{Subedi:2014ik,Mankowsky:2014vt,Fechner:2016tu}. For a single optical pulse, this structural distortion decays and the system relaxes back to the ground state, whereas continous driving yields a combination of time-dependent and time-independent structural distortions. We will discuss these distortions later based on the analytical solution of Eqn.~\eqref{eq_anharmonic}.


To model the magnetic structure we consider a Heisenberg Hamiltonian with
\begin{equation}\label{eq_Heisenberg}
	H^{\mathrm{mag}}=\sum\limits_{\braket{i,j}} J_{i,j}(S_i\cdot S_j) + D\sum\limits_{i=1}^N (S_i^z)^2\;,
\end{equation}
where $S_i$ is a localized spin magnetic moment, $J_{i,j}$ are the magnetic exchange interactions between spins $i$ and $j$, and $D$ is the uniaxial magneto-crystalline anisotropy (MCA) energy. We introduce the coupling of the local spin moments contained in the Heisenberg Hamiltonian, to the distortion $\xi$ of the lattice Hamiltonian by expanding the magnetic exchange interactions with respect to the distortion\cite{Wang:2008dfa,Fechner:2016tu}. For an expansion up to second order we obtain the following spin-phonon coupling Hamiltonian:
\begin{equation}\label{eq_spin_phonon_general}
	H^{\mathrm{sp}}=\sum\limits_{\braket{i,j}} \frac{\partial J_{i,j}}{\partial \xi}(S_i\cdot S_j)\xi +\sum\limits_{i,j} \frac{\partial^2 J_{i,j}}{\partial \xi^2}(S_i\cdot S_j)\xi^2 \quad .
\end{equation}
Note that the first derivatives of exchange with respect to mode $\xi$ can be zero for certain mode symmetries, and in general only the second order derivatives are non-zero. For the symmetry-conserving Raman modes, $\xi_\mathrm{R}$, however, the first order spin-lattice coupling is non-zero; note also that these are the modes that have a quadratic-linear lattice coupling with the IR modes in Eqn.~\eqref{eq_anharmonic}. Since the $\xi_\mathrm{R}$ distortion is symmetry conserving we can directly write the exchange interaction as a function of \rev{the}  mode amplitude as
\begin{equation}\label{eq_single_exchange}
	J_{i,j}(\xi_{\mathrm{R}})=J_{i,j}+ \frac{\partial J_{i,j}}{\partial \xi_{\mathrm{R}}} \xi_{\mathrm{R}} + \frac{\partial^2 J_{i,j}}{\partial \xi^2_{\mathrm{R}}} \xi^2_{\mathrm{R}} + \dots  \quad,
\end{equation}
with the same labeling as in Eqn.~\eqref{eq_Heisenberg}. \rev{(For a phonon mode of general symmetry, either Raman or IR active, the situation is more complicated since the symmetry breaking can split degenerate exchange interactions, resulting in an increased total number of inequivalent exchange interaction parameters).} \rev{In principle the MCA energy term, $D$, is also a function of the mode amplitude. However, we find that its variation is negligible for \CrO.}

\subsection{Computational details}

To calculate the structure, phonons and magnetic exchange interactions of \CrO\  we use density functional theory with the local spin density approximation plus Hubbard $U$ (LSDA$+U$) exchange-correlation functional. We use parameters $U$=\unit[4]{eV} and $J$=\unit[0.5]{eV} on the Cr-$d$ orbitals and treat the double counting correction within the fully-localized limit. These parameters have been shown to give a good description of \CrO\ in earlier work\cite{Shi:2009jka,Mostovoy:2010ia,Mu:2014gp}. We use the Vienna \textit{ab-initio} simulation package (VASP) \cite{Kresse:1996vf} within the projector augmented wave (PAW) method \cite{Blochl:1994uk} using default VASP PAW pseudopotentials generated with the following valence-electron configurations: Cr ($3s^2 3p^6 4s^1 3d^{5}$), O ($2s^23p^4$). We sample the Brillouin zone in our total energy calculations using 11$\times$11$\times$11 and 9$\times$9$\times$5 $k$-point meshes for the primitive rhombohedral and hexagonal cells respectively, and use a plane-wave energy cutoff of \unit[600]{eV}. Finally, for computing the MCA energy of \CrO\ we use an increased $k$-point grid of 14$\times$14$\times$14 within the rhombohedral cell.

Previous theoretical studies of \CrO\  have addressed the microscopic origin of the magnetoelectric effect\cite{Iniguez:2008gm,Mostovoy:2010ia,Malashevich:2012et} and the magnetic properties\cite{Shi:2009jka,Mu:2014gp} using a combination of first-principles density functional theory (DFT) calculations and effective Hamiltonian approaches. These studies demonstrated that magnetoelectric properties, phonon frequencies and magnetic exchange interactions, all key quantities in this work, are well described by DFT calculations with technical details similar to those chosen here. 

We \rev{calculate the atomistic} spin-dynamics by solving the Landau-Lifshitz Gilbert equation numerically using the Heun method\cite{Sukhov:2008cz} with an integration time step that is one thousandth of the fasted period of the oscillations $(\dfrac{\pi\mu}{\gamma D}10^{-3}\approx 4 fs)$.

\section{C\lowercase{r}$_2$O$_3$}
\label{Cr2O3}

\CrO\  crystallizes in the corundum structure which is composed of a combination of edge- and face-sharing CrO$_6$ octahedra. The magnitude 3$\mu_B$ spin magnetic moments on the $d^4$ Cr$^{3+}$ ions order antiferromagnetically below the N\'eel temperature, $T_\mathrm{N}=$\unit[307]{K}, in a collinear ``$(\uparrow,\downarrow,\uparrow,\downarrow)$'' pattern with magnetic space group $R\overline{3}'c$ (161) that breaks inversion symmetry  (Fig.~\ref{fig_1})\cite{Foner:1963vi,Stone:1971gj}. The primitive unit cell, with its four chromium and six oxygen atoms is shown in Fig.~\ref{fig_1}(a). As a result of its simultaneous breaking of time-reversal and space-inversion symmetry, \CrO\  exhibits the linear magnetoelectric effect, in which a magnetic/electric field induces an electric/magnetic polarization. Indeed, \CrO\  is considered to be the prototypical magnetoelectric, being the material in which the effect was first predicted \cite{Dzyaloshinskii:1960} and subsequently measured \cite{Astrov:1960vt}.

\subsection{Calculated lattice properties of \CrO}

We begin by calculating the lowest-energy structure of \CrO\  by relaxing its rhombohedral unit cell to obtain a force-free DFT reference structure. We initialized our computations using data from the experimental study of Ref.~[\onlinecite{Finger:1980kx}] and optimized the structure until the forces on each atom were less than \unit[0.01]{meV/\AA}. The resulting structure has a unit cell volume of \unit[96.46]{\AA$^3$}, with the coordinates $x=0.152$ for Cr and $x=0.304$ for O at the Wyckoff positions $4c$ and $6e$, respectively, in good agreement with literature experimental\cite{Finger:1980kx} and theoretical\cite{Mostovoy:2010ia} values.

\begin{figure}[bt]
\includegraphics[width=86mm]{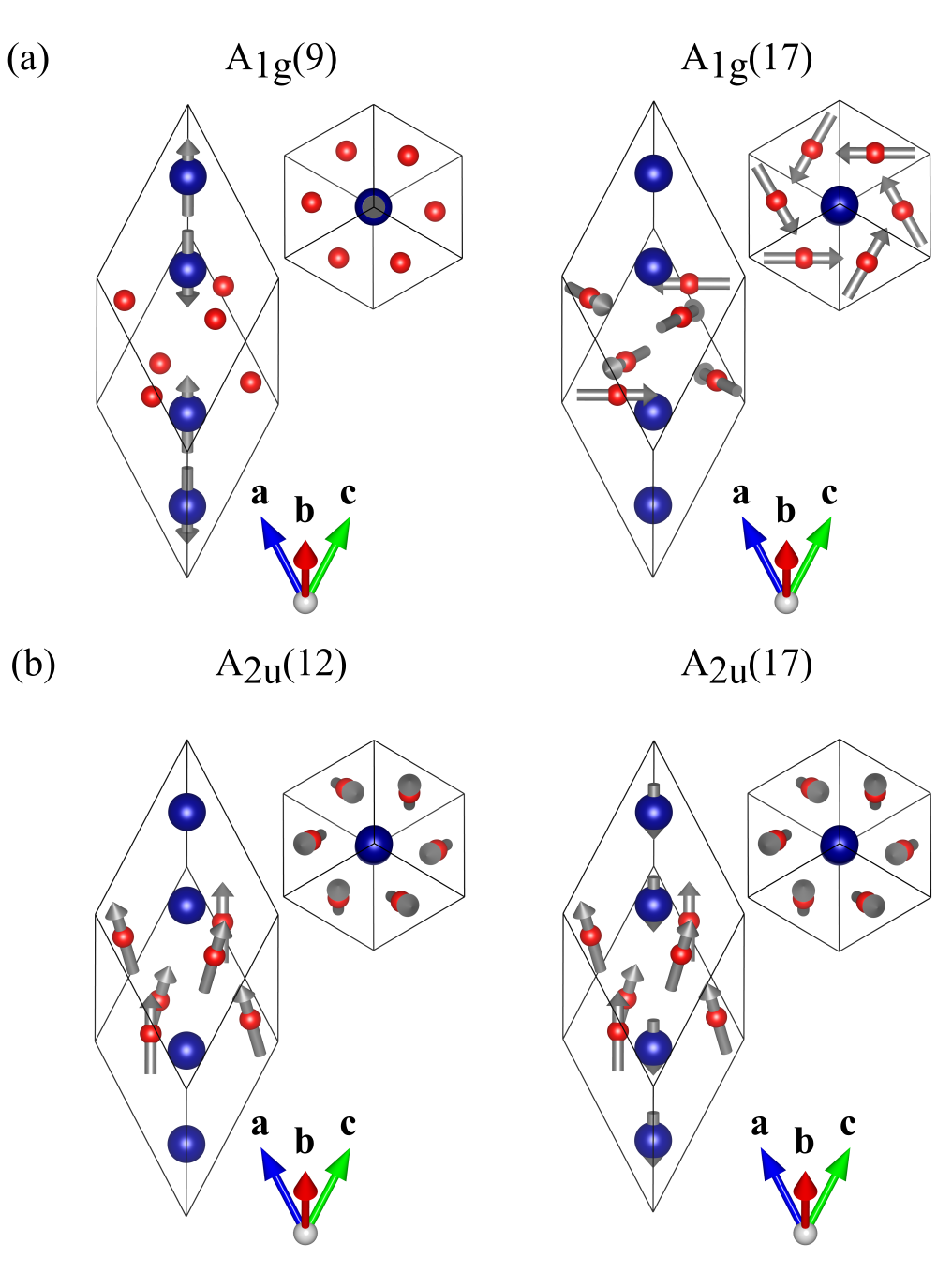}
\caption{\label{fig_2} Displacement pattern of the \CrO\  phonon modes relevant in this work. (a) shows symmetry-conserving \PH{A}{1g} modes, and (b) displays IR-active \PH{A}{2u} modes. The grey arrows show the displacement direction of each atom for the specific mode, with the indicated directions defining positive displacement amplitudes. The notation indicates the irreducible representation for the mode symmetry followed in brackets by the calculated mode frequency in THz, rounded to the nearest integer.}
\end{figure}

\begin{table}[bt]
\caption{\label{tab_1} Phonon frequencies of symmetry conserving Raman and infrared-active modes of \CrO\ in THz. The experimental values (EXP) are taken from Refs.~[\onlinecite{Beattie:1970ks,Lucovsky:1977uu,Shim:2004eo}]. The displacement patterns of the \PH{A}{1g} and \PH{A}{2u} modes are shown in Fig.~\ref{fig_2} (a,b).}
\begin{ruledtabular}
\begin{tabular}{|r|r|r|}sym.	&DFT	 			&EXP	 \rule[-1ex]{0pt}{3.5ex}\\\hline
$A_{1g}$&9.3			 	&9.0	 \rule[-1ex]{0pt}{3.5ex}\\
$A_{1g}$&17.3			&16.5	 \rule[-1ex]{0pt}{3.5ex}\\ 
$A_{2g}$&8.0			 	&--	 \rule[-1ex]{0pt}{3.5ex}\\
$A_{2g}$&13.8			&--	 \rule[-1ex]{0pt}{3.5ex}\\ 
$A_{2g}$&20.7		 	&--	 \rule[-1ex]{0pt}{3.5ex}\\
$E_{g}$&9.2			 	&8.7	 \rule[-1ex]{0pt}{3.5ex}\\
$E_{g}$&10.7				&10.5 \rule[-1ex]{0pt}{3.5ex}\\ 
$E_{g}$&12.4				&12.0	 \rule[-1ex]{0pt}{3.5ex}\\ 
$E_{g}$&16.1				&15.6 \rule[-1ex]{0pt}{3.5ex}\\ 
$E_{g}$&19.2				&18.5 \rule[-1ex]{0pt}{3.5ex}\\ 
$A_{2u}$&12.2			&12.1	 \rule[-1ex]{0pt}{3.5ex}\\
$A_{2u}$&17.2			&16.0	 \rule[-1ex]{0pt}{3.5ex}\\
$E_{u}$	&9.3 			&9.1	 \rule[-1ex]{0pt}{3.5ex}\\
$E_{u}$	&13.5			&13.2	 \rule[-1ex]{0pt}{3.5ex}\\
$E_{u}$	&17.0			&16.1	 \rule[-1ex]{0pt}{3.5ex}\\
$E_{u}$	&19.0			&18.2	 \rule[-1ex]{0pt}{3.5ex}\\
\end{tabular}
\end{ruledtabular}

\end{table}

Next, we compute the phonon frequencies and eigenvectors of our ground-state structure using density functional perturbation theory \cite{Baroni:2001tn}. Light radiation only excites polar phonon modes close to the center of the Brillouin zone, $q=(0,0,0)$. Consequently, we do not calculate the full phonon band structure but only the modes at this special point in reciprocal space. Since the primitive cell of \CrO\ contains 10 atoms, there are 27 non-translational zone-center phonon modes, which span the irreducible representatives of the $\overline{3}'m$ point group: $2 A_{1g} \otimes 3 A_{2g}\otimes 2 A_{1u}\otimes 2 A_{2u} \otimes 10 E_{g} \otimes 8 E_{u}$. Of these modes only the \PH{A}{2u} and \PH{E}{u} modes are polar, with the dipole moments of the \PH{A}{2u} modes pointing along the long rhombohedral axis ($a$+$b$+$c$) and those of the \PH{E}{u} modes perpendicular to it. The \PH{A}{1g} modes, which are not directly excitable by light, have the symmetry of the \CrO\ point group and consequently exhibit a square-linear coupling to the polar modes in the anharmonic potential. We list in Tab.~\ref{tab_1} the computed frequencies of \PH{A}{1g} and optically active modes together with available experimental frequencies from the literature\cite{Beattie:1970ks,Lucovsky:1977uu,Shim:2004eo}, and find good agreement. 

In Fig.~\ref{fig_2} we show the displacement patterns of the \PH{A}{2u} and \PH{A}{1g} modes with the grey arrows indicating the direction of displacement of the atoms for positive mode amplitude. Within the \CrO\ structure the \unit[9.3]{Thz} \PH{A}{1g} (\PH{A}{1g}(9)) mode modulates the Cr-Cr distance, whereas the higher frequency \unit[17.3]{Thz} \PH{A}{1g}(17) mode modulates the Cr-O-Cr bond-angles via a rotation of the oxygen octahedra around the rhombohedral axis. Both polar \PH{A}{2u} modes exhibit a collective motion of the oxygens along the rhombohedral axis, with the \unit[17.2]{Thz} \PH{A}{2u}(17) mode involving the larger relative movement of the Cr and oxygen atoms. The Cr-Cr bond lengths are unchanged by the movement patterns of the polar modes. 

\begin{figure}[bt]
\includegraphics[width=86mm]{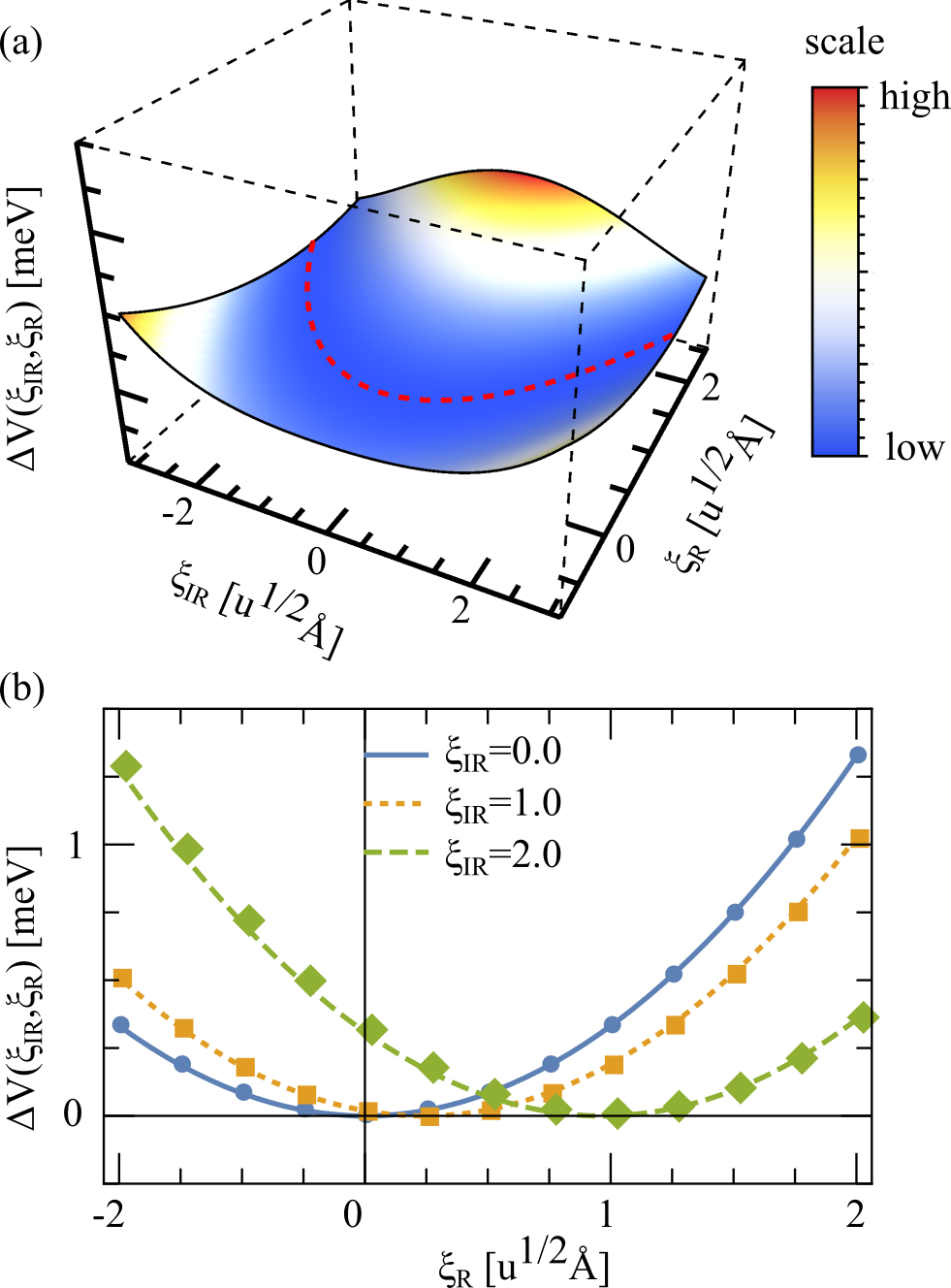}
\caption{\label{fig_3} (a) Calculated two-dimensional potential surface of the anharmonic phonon-phonon interaction between $\xi_\mathrm{R}=$\PH{A}{1g}(9) and $\xi_\mathrm{IR}=$\PH{A}{2u}(17). The red line in the three dimensional plot shows the position of the potential minimum. (b) selected cuts through the two dimensional potential surface shown in (a). Note that we plot $\Delta V(\xi_\mathrm{IR},\xi_\mathrm{R})=V(\xi_\mathrm{IR},\xi_\mathrm{R})-V(\xi_\mathrm{IR},-g\xi^2_\mathrm{IR}/(2\oR^2))$, so that the minimum is set to \unit[0]{meV}.}
\end{figure}
With our calculated phonon eigenvectors as the starting point, we next compute the anharmonic phonon coupling constants by mapping the potential of Eqn.~\eqref{eq_anharmonic} onto total energy calculations of \CrO\ structures, distorted by appropriate superpositions of the phonon eigenvectors as in previous work\cite{Fechner:2016tu,Juraschek:2017ic}. We are primarily interested in the quadratic-linear coupling of Eqn.~\eqref{eq_anharmonic}, which is only nonzero if the linear component has the full point group symmetry, which is \PH{A}{1g} for \CrO. For convenience, we assume that the radiation is oriented along the rhombohedral axis such that only \PH{A}{2u} modes are directly excited, then we compute the 2D-potential of Eqn.~\eqref{eq_anharmonic} for all combinations of polar \PH{A}{2u} and \PH{A}{1g} modes.

In Fig.~\ref{fig_3} (a) we show the computed potential landscape for the combination of the \PH{A}{1g}(9) and \PH{A}{2u}(17) phonon modes. Displacement of the \PH{A}{2u}(17) mode causes a shift of the potential minimum of the \PH{A}{1g}(9) mode, as shown in the cuts of the 2D-potential in Fig.~\ref{fig_3} (b). The red dashed line in Fig.~\ref{fig_3} (a) shows the position of the \PH{A}{1g}(9) mode minimum within the 2D potential landscape. For negative and positive amplitudes of the \PH{A}{2u}(17) mode, the potential minimum position shifts to positive amplitudes of the \PH{A}{1g}(9), corresponding to a negative sign of the square-linear coupling. We quantify this observation by fitting the complete potential landscape using Eqn.~\eqref{eq_anharmonic}, to extract all anharmonic coupling constants and repeat the calculation for all combinations of \PH{A}{2u} and \PH{A}{1g} modes. The  computed anharmonic constants are given in Tab.~\ref{tab_2}.

\begin{table}[bt]
\caption{\label{tab_2} Upper panel: Anharmonic coupling constants $g$, in units of [meV/($\sqrt{u}$\AA)$^3$], between symmetry conserving \PH{A}{1g} and IR active phonon modes of \PH{A}{2u} symmetry. Lower panel: quartic anharmonic constants, $\gamma$, in units of [meV/($\sqrt{u}$\AA)$^4$].}
\begin{ruledtabular}
\begin{tabular}{|c|r|r|}
modes&\PH{A}{2u}(12)&  \PH{A}{2u}(17)\rule[-1ex]{0pt}{3.5ex}\\\hline
\PH{A}{1g}(9)	&6&-86\rule[-1ex]{0pt}{3.5ex}\\
\PH{A}{1g}(17)	&-38&101\rule[-1ex]{0pt}{3.5ex}\\
\end{tabular}
\begin{tabular}{|c|r|r|r|r|}
modes&\PH{A}{1g}(9)&\PH{A}{1g}(17)&\PH{A}{2u}(12)&\PH{A}{2u}(17)\rule[-1ex]{0pt}{3.5ex}\\\hline
$\gamma_\mathrm{IR}$	&1&4&4&14\rule[-1ex]{0pt}{3.5ex}\\\hline
\end{tabular}\end{ruledtabular}
\end{table}

We find that the nominal value of the quadratic-linear anharmonic coupling $g$ varies from 6 to \unit[101]{meV/($\sqrt{u}$\AA$)^3$} and exhibits positive or negative sign, so that modulations of the \CrO\ structure with positive and negative amplitudes of the \PH{A}{1g} modes can be induced by exciting the appropriate polar mode. (Note that the opposite choice of sign in the definition of the phonon eigenvectors would reverse the sign of $g$; the signs given in Tab.~\ref{tab_2} correspond to the phonons as defined in Fig.~\ref{fig_2}). 

Minimization of Eqn.~\eqref{eq_anharmonic} gives the amount of induced structural distortion to be $\xi_{\mathrm{R}}\approx -g\,\xi^2_{\mathrm{IR}}/\omega^2_{\mathrm{R}}$. Consequently, for the combination of polar \PH{A}{2u}(12) and \PH{A}{1g}(9) modes, excitation of the polar mode induces, due to the positive coupling constant $g$, a negative amplitude of the \PH{A}{1g}(9) mode which results in a decrease in the nearest-neighbor Cr-Cr distance. In contrast, the \PH{A}{2u}(17) mode couples with a negative coupling constant $g$ to the \PH{A}{1g}(9) mode and so the induced quasi-equilibrium structure has an increased Cr-Cr distance.  The \PH{A}{1g}(17) mode changes the oxygen octahedral rotation angles around the Cr ions. Its negative coupling to the \PH{A}{2u}(12) mode results in a decreased rotational angle, whereas the positive coupling to the \PH{A}{2u}(17) mode increases the rotational angle in the quasi-equilibrium structure.

\subsection{Calculated magnetic properties of \CrO}

The fact that the transient structure generated through the quadratic-linear coupling of the optically excited polar modes to the \PH{A}{1g}(9) Raman mode has a modified Cr-Cr distance suggests that it might also have a different magnetic ground state. To explore this possibility, we next calculate the energy difference between the AFM ground-state ordering $(\uparrow,\downarrow,\uparrow,\downarrow)$ and \rev{two other magnetic orderings of the Cr spins -- ferromagnetic (FM) $(\uparrow,\uparrow,\uparrow,\uparrow)$ and another antiferromagnetic (AFM$_\mathrm{1}$) $(\uparrow,\uparrow,\downarrow,\downarrow)$ -- as a function of the \PH{A}{1g}(9) distortion amplitude.}
 
\rev{For the equilibrium structure, we find that the AFM$_\mathrm{1}$ state is \unit[67]{meV} and the FM state \unit[162]{meV} in energy above the AFM ground state. Modulating the structure with the pattern of atomic displacements corresponding to the \PH{A}{1g}(9) phonon mode in the positive direction, so that the Cr-Cr nearest-neighbor distance, ($d_{\mathrm{Cr-Cr}}$), is increased, significantly lowers both of these energy differences. For positive amplitudes larger than $\xi_{\mathrm{R}}\geq$\unit[0.75]{$\sqrt{u}$\AA}, corresponding to a stretching of $d_{\mathrm{Cr-Cr}}=$\unit[0.06]{\AA}, the energy of the AFM$_\mathrm{1}$ state becomes lower than the AFM ground state; at larger amplitudes ($\xi_{\mathrm{R}}\geq$\unit[1.9]{$\sqrt{u}$\AA}) the FM state becomes lower in energy than the original ground state, but remains higher in energy than the AFM$_\mathrm{1}$ state. We therefore predict that a crossover to the AFM$_\mathrm{1}$ state should be achievable through quadratic-linear coupling with appropriate choice of the polar mode excitation frequency and intensity. (Note that modulating the structure with a negative amplitude of \PH{A}{1g}(9), which decreases the Cr-Cr nearest-neighbor bond, increases the relative energies of the FM and AFM$_\mathrm{1}$ states). In contrast, modulating the structure along the eigenvector of the second \PH{A}{1g} mode at \unit[17]{THz}, or along those of the polar \PH{A}{2u} modes has only a small effect on the magnetic energy landscape.}

\begin{figure*}[tb]
\includegraphics[width=172mm]{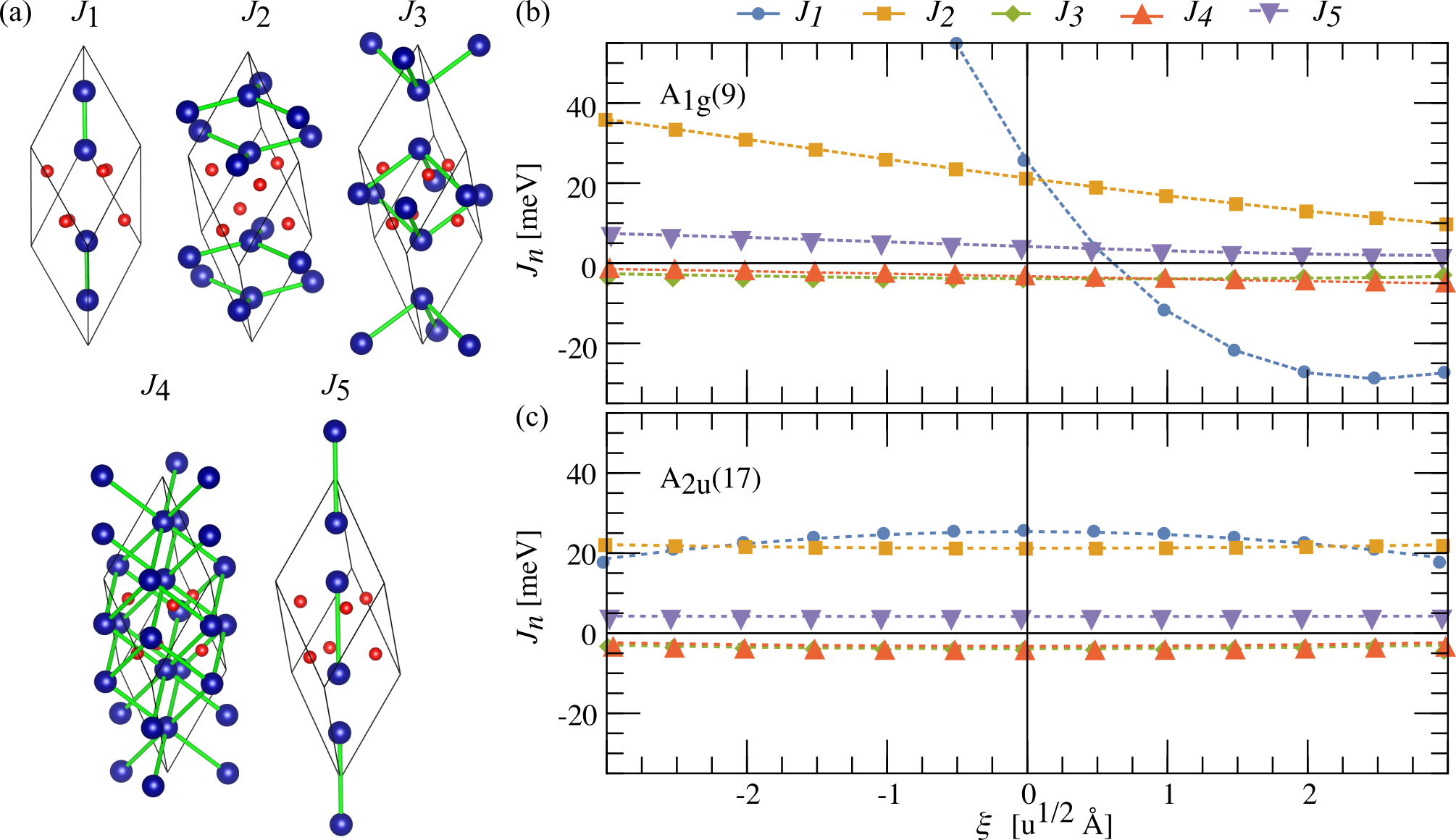}
\caption{\label{fig_4} (a) Illustration of the magnetic exchange interactions in \CrO, from first to fifth nearest neighbor. (b,c) Changes in the magnetic exchange interactions due to structural modifications by the \PH{A}{1g}(9) and \PH{A}{2u}(17) modes. Note that for the \PH{A}{1g}(9) mode the nearest-neighbor magnetic exchange ($J_1$) changes sign for negative phonon mode amplitudes.}
\end{figure*}

To explore the magnetic behavior further, we next calculate the magnetic exchange interactions of the ground-state structure using the Heisenberg Hamiltonian of \eqref{eq_Heisenberg}, including magnetic exchange interactions, $J_{n}$, up to fifth nearest neighbors, as shown in Fig.~\ref{fig_4}; this Hamiltonian has been shown to give an accurate theoretical description of the magnetoelectric effect and magnetic transition temperature of \CrO\ \cite{Shi:2009jka,Mostovoy:2010ia}. \rev{Specifically, our Heisenberg Hamiltonian for the magnetic exchanges reads:}
\begin{align}
	H^\mathrm{exch}_\mathrm{Cr_{2}O_{3}}=&\; J_1(S_1 \cdot S_2 + S_3 \cdot S_4 )\nonumber \\
	&+3J_2(S_1 \cdot S_4 + S_2 \cdot S_3 )\nonumber \\
	&+3J_3(S_1 \cdot S_2 + S_3 \cdot S_4 )\label{eq_cr2o3exchanges} \\
	&+6J_4(S_1 \cdot S_3 + S_2 \cdot S_4 )\nonumber \\
	&+J_5(S_2 \cdot S_3 + S_1 \cdot S_4 )\nonumber \;\;,
\end{align}
\rev{with the $J_n$ as shown in Fig.~\ref{fig_4}, and the labeling of spins as in} \rev{Fig.˜\ref{fig_1}}. We extract the magnetic exchange interactions from the total energy differences between four distinct magnetic arrangements within the non-primitive hexagonal cell, using the approach of Ref.~[\onlinecite{Xiang:2011cn}]. The resulting magnetic exchange interactions are listed in Tab.~\ref{tab_3} and are in agreement with earlier theoretical works \cite{Shi:2009jka,Mostovoy:2010ia,Mu:2014gp}. We find the nearest and next-nearest neighbor interactions, $J_1$ and $J_2$, to be strongly antiferromagnetic, whereas $J_3$ and $J_4$ favor ferromagnetic arrangements. The furthermost exchange interaction that we consider, $J_5$, is weakly antiferromagnetic. Finally, we note that, in contrast to other magnetic insulators\cite{Fedorova:2015fu}, higher-order magnetic exchanges such as four-body interactions are not required for the description of the magnetism in \CrO\ \cite{Mu:2014gp}.

Next, we compute how the modulation of the \CrO\ structure by the phonon mode eigenvectors changes the magnetic exchange interactions, using the same approach to extract the exchange interactions as we used above for the ground-state structure. \rev{(For the \PH{A}{2u} modes we neglect the small splittings in $J$ values that result from the lowered symmetry.)} Our calculated coefficients of the expansion of Eqn.~\eqref{eq_single_exchange}, listed up to quadratic order in $\xi$ in Table~\ref{tab_3}, are a measure of the spin-phonon coupling for each mode. In Fig.~\ref{fig_4} (b,c), we plot the five nearest-neighbor magnetic exchange constants as a function of the \PH{A}{1g}(9) and \PH{A}{2u}(17) phonon mode amplitudes. We find that the \PH{A}{1g}(9) mode significantly changes the nearest-neighbor exchange interaction, whereas the longer range magnetic exchange interactions are less affected by the structural modulation. \rev{An intriguing result is the sign change of the nearest-neighbor exchange interaction $J_1$ at amplitudes $\xi\geq$\unit[0.75]{$\sqrt{u}$\AA}, corresponding to an increase of \unit[0.06]{\AA} in the Cr-Cr bond length, consistent with the crossover to AFM$_1$} ordering that we found above. In contrast to the \PH{A}{1g}(9) mode we see that the \PH{A}{2u}(17) mode has minimal direct effect on the magnetic exchange interactions. The other \PH{A}{1g} and \PH{A}{2u} modes (not shown) also have minimal effect on the exchange interactions. The spin-phonon coupling constants obtained by fitting these results to Eqn.~(\ref{eq_single_exchange}) are listed in Tab.~\ref{tab_3}; as expected the coefficients of $J_1$ for the \PH{A}{1g}(9) mode are large. 

\begin{table}[bt]
\caption{\label{tab_3} Upper panel: Magnetic exchange interactions (meV) for the ground-state structure of \CrO. Lower panel: Spin-phonon coupling constants (units meV/($\sqrt{u}$\AA) and meV/($\sqrt{u}$\AA)$^2$ for first/second order) for the \PH{A}{1g} and \PH{A}{2u} modes of \CrO. }
\begin{ruledtabular}
\begin{tabular}{|r|r|r|r|r|}
$ J_1$ 	& $ J_2$ 	& $ J_3$ & $ J_4$ 	& $ J_5$\rule[-1ex]{0pt}{3.5ex}	\\\hline
 25.4 	& 21.2 		& -3.9 	 & -3.3 	& 4.2 	\rule[-1ex]{0pt}{3.5ex} \\
\end{tabular}
\begin{tabular}{|c|r|r|r|r|r|}
n									&1  	& 2 		& 3 	& 4 	& 5     	\rule[-1ex]{0pt}{3.5ex}	\\\hline
\multicolumn{6}{|l|}{\PH{A}{1g}(9)\rule[-1ex]{0pt}{3.5ex}}\\\hline
$\partial J_n/\partial\xi$			& -57.9 &-4.4 	& -0.1 & -0.6 & -1.0 			\rule[-1ex]{0pt}{3.5ex} \\
$\partial^2 J_{n}/\partial\xi^2$	& 14.5 &0.2 	& 0.1 & 0.0 &  0.1 			\rule[-1ex]{0pt}{3.5ex} \\\hline
\multicolumn{6}{|l|}{\PH{A}{1g}(17)\rule[-1ex]{0pt}{3.5ex}}\\\hline
$\partial J_n/\partial\xi$			&12.4 &1.4 	& -0.0& 0.1 & 0.3 			\rule[-1ex]{0pt}{3.5ex} \\
$\partial^2 J_{n}/\partial\xi^2$	& 3.9 &0.4 	& 0.1 & 0.0 &  0.0 			\rule[-1ex]{0pt}{3.5ex} \\\hline
\multicolumn{6}{|l|}{\PH{A}{2u}(12)\rule[-1ex]{0pt}{3.5ex}}\\\hline
$\partial^2 J_{n}/\partial\xi^2$	& 0.5 &0.1 	& -0.1 & 0.0 &  0.2 			\rule[-1ex]{0pt}{3.5ex} \\\hline
\multicolumn{6}{|l|}{\PH{A}{2u}(17)\rule[-1ex]{0pt}{3.5ex}}\\\hline
$\partial^2 J_{n}/\partial\xi^2$	& -0.7 &0.1 	& 0.0 & 0.0 &  0.0 			\rule[-1ex]{0pt}{3.5ex} \\
\end{tabular}
\end{ruledtabular}
\end{table}

We can understand the strong $J_1$ response by analyzing the displacement pattern of the \PH{A}{1g}(9) mode in the context of the origin of the $J_1$ magnetic exchange interaction that has been discussed in the literature. Earlier analysis of the magnetic interactions in the ground-state of \CrO\ \cite{Shi:2009jka} showed that the main contribution to $J_1$ arises from an antiferromagnetic direct exchange interaction between the nearest Cr atoms combined with a small ferromagnetic superexchange component from the \unit[82]{$^\circ$} Cr-O-Cr interaction. For positive amplitudes of the \PH{A}{1g}(9) mode, the Cr-Cr distance increases thus decreasing the antiferromagnetic direct exchange interaction. At the same time, the Cr-O-Cr angle becomes closer to \unit[90]{$^\circ$} enhancing the ferromagnetic superexchange. The result is a change in sign of $J_1$. We note that this observation is possibly connected to the findings of Ref.~[\onlinecite{Shi:2009jka}], in which strong modulations of magnetic energies induced by small changes of the \CrO\  ground-state structure were reported. Moreover, since the direct magnetic exchange interaction only affects $J_1$, the magnetic exchange interactions $J_n$ with $n\ge2$ are less affected by the structural distortion. 

Finally, we calculate the MCA energy of \CrO, from the energy difference between alignment of the Cr spin moments along ($E_{||}$) and perpendicular ($E_{\perp}$) to the rhombohedral axis, including the spin-orbit interaction in our calculations. We obtain an energy difference $E_{||}-E_{\perp}$= \unit[-27]{$\mu$eV}; the experimental\cite{Foner:1963vi,Dudko:1971fv,Tobia:2010dl} values range from \unit[-12]{$\mu$eV} to  \unit[-16]{$\mu$eV}. We also calculate the change in MCA energy when the structure is modulated by the \PH{A}{1g} or \PH{A}{2u} phonon modes and find no significant change (a mode amplitude of $\xi=$\unit[$\pm$2]{$\sqrt{u}$\AA} lowers the MCA energy by \unit[$\approx$10]{\%}). In particular, the rhombohedral easy axis is preserved upon structural modulation. This finding justifies our omission of MCA terms in our spin-phonon Hamiltonian, Eqn.~\eqref{eq_spin_phonon_general}. 

To summarize this section, we find a strong dependence of the $J_1$ nearest-neighbor magnetic exchange interaction on the structural distortion associated with the \PH{A}{1g}(9) mode, with positive mode amplitude, corresponding to increased Cr-Cr distance, inducing a change in sign. This dependence leads to a crossover \rev{between antiferromagnetic states}. Since the \PH{A}{1g}(9) mode couples quadratic-linearly to the \PH{A}{2u} modes, this crossover can be induced by optical excitation of the polar modes. Following the classical considerations derived in Refs.~[\onlinecite{Fechner:2016tu,Juraschek:2017ic}], we estimate that a pulse fluence of $\sim$\unit[40]{mJ/cm$^2$} at a frequency \unit[17]{THz} should be sufficient to induce this crossover transition. \rev{A similar fluence was reported  in Ref.~[\onlinecite{Mankowsky:2017ur}] without damaging the sample.}

\section{NON-linear lattice dynamics}
\label{NLLD}

Having established that the structural modification induced via non-linear phononic coupling can lead to a change in magnetic ordering, we next evaluate the dynamical behavior associated with driving a phonon. We begin by calculating the non-linear lattice dynamics using the vibrational crystal potential given in Eqn. \eqref{eq_anharmonic}, followed by the resulting spin dynamics. We study the case in which an IR mode is excited by a sinusoidal driving force $F(t)$ with amplitude $E_\mathrm{drive}$ with frequency $\Omega$ and calculate the resulting dynamics of the coupled R mode, focussing particularly on the combination of the \PH{A}{2u}(17) and \PH{A}{1g}(9) which yields a negative amplitude \PH{A}{1g}(9) displacement and possible ferromagnetism. The time evolution of the system described by the potential of Eqn.~\eqref{eq_anharmonic} is then governed by the following set of differential equations:
\begin{eqnarray}\label{eq_dynamics_phonon}
\displaystyle \ddot{\xi}_{\mathrm{IR}} +\omega^2_{\mathrm{IR}}\xi_{\mathrm{IR}}+ \gamma_\mathrm{IR} \xi^3_{\mathrm{IR}} &=& 2g \xi_{\mathrm{IR}}\xi_{\mathrm{R}}  + F(t), \\
\displaystyle \ddot{\xi}_{\mathrm{R}} +\omega^2_{\mathrm{R}}\xi_{\mathrm{R}}+\gamma_\mathrm{R} \xi^3_{\mathrm{R}} &=& g \xi^2_{\mathrm{IR}}, \\
\displaystyle F(t) &=&E_\mathrm{drive} \sin(\Omega t)\;.
\end{eqnarray}
We derive a closed analytical solution of the dynamic equations in the limit in which the coupling and the anharmonicity are small relative to the frequency, that is $\frac{g\;\gamma_\mathrm{IR}\;\gamma_\mathrm{R}}{\omega_{\mathrm{IR}}\;\omega_{\mathrm{R}}}\ll 1$ by following the approach of Ref.~[\onlinecite{Lichtenberg:1992vb}]. For the case of \CrO, our {\it ab initio} values, provided in Table \ref{tab_1} and \ref{tab_2}, indicate that the combination of \PH{A}{1g} with polar \PH{A}{2u} modes fulfills this criterion. The dynamics of the IR mode are then given by: 
\begin{eqnarray}\label{eq_IRdynamics}
\xi_\mathrm{IR}(t)&=&\AIR \sin[\noIR t]+\AOM \sin[\Omega t]\nonumber\\
&&+\frac{g \AIR \AR}{2[\oIR^2-(\oIR-\oR)^2]} \cos[(\noIR-\noR)t]\nonumber\\
&&+\frac{g \AIR \AR}{2[\oIR^2-(\oIR+\oR)^2]} \cos[(\noIR+\noR)t]\\
&&+\frac{g \AOM \AR}{2[\oIR^2-(\Omega-\oR)^2]} \sin[(\Omega-\noR)t]\nonumber\\
&&+\frac{g \AOM \AR}{2[\oIR^2-(\Omega+\oR)^2]} \sin[(\Omega+\noR)t]\nonumber\;,
\end{eqnarray}
and those of the R mode by:
\begin{eqnarray}\label{eq_Rdynamics}
\xi_\mathrm{R}(t)&=&\xi_\mathrm{R0}+\AR \cos[\noR t] \nonumber\\
&&+\frac{g \AIR^2}{4 [\oR^2-4\oIR^2]} \cos[2\noIR t]\nonumber\\
&&+\frac{g \AOM^2}{4 [\oR^2-4\Omega^2]} \cos[2\Omega t]\\
&&+\frac{g\AIR \AOM}{2 [\oR^2-(\Omega+\oIR)^2]} \sin[(\Omega+\noIR) t]\nonumber\\
&&+\frac{g\AIR \AOM}{2 [\oR^2-(\Omega-\oIR)^2]} \sin[(\Omega-\noIR) t]\nonumber\;.
\end{eqnarray}
The time-independent displacement of the R mode oscillation is given by $\xi_0=g(\AIR^2+\AOM^2)/(4\oR^2)$, with the amplitude factors, $\AIR$ and $\AR$ depending on the initial amplitudes, $\xi_\mathrm{R}(0)$ and $\xi_\mathrm{IR}(0)$ and $\AOM=1/(\oIR^2-\Omega^2)$. We indicate frequencies with a tilde which have been renormalized by the anharmonic coupling, as given by Eqns.~\eqref{eqn_correctionomegaIR} and \eqref{eqn_correctionomegaR} in the Appendix.

The solution shows that the anharmonic potential and the coupling between the phonon modes induce motions of the oscillators which display several components given by cosine and sine terms. Each of these terms corresponds to a single component of the motion with a specific amplitude and frequency -- either the renormalized original frequency of each oscillator, indicated by the tilde, or sums or differences of the original frequencies. We emphasize that these motions arise from a single mode, which exhibits multiple frequencies because of its anharmonicity and coupling.

\begin{figure}[bt]
   \centering
   \includegraphics[width=1\columnwidth]{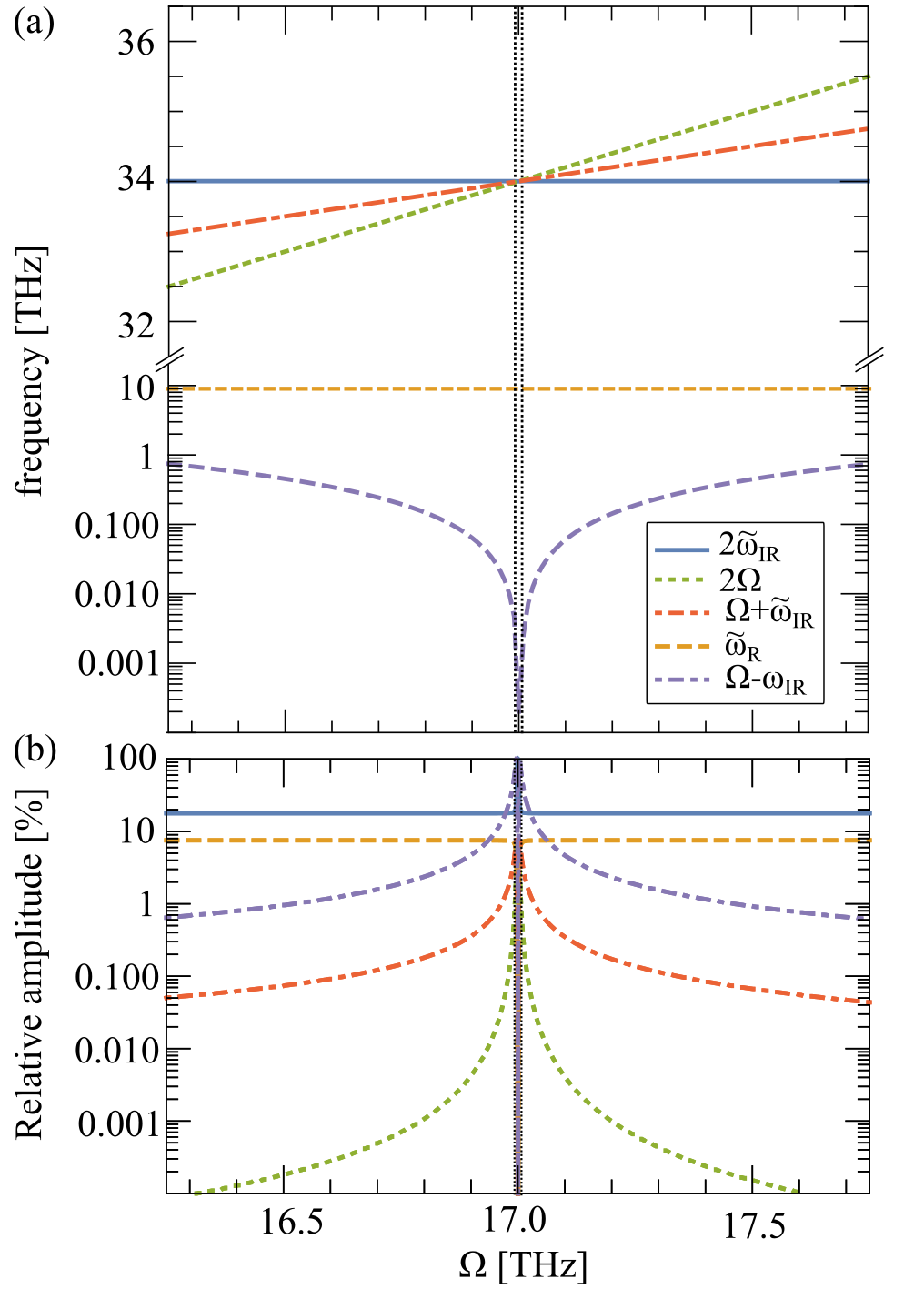}
   \caption{\label{fig_5} (a) Frequencies, $\omega$, and (b) relative amplitudes, $\xi$ (normalized to $\xi_\mathrm{R0}$), of the five separate parts of the R mode motion as a function of the external driving frequency. Note the separation and the linear and logarithmic scales in (a).}
\end{figure}

Next we analyze the frequencies and amplitudes of each term in Eqn.~\ref{eq_Rdynamics} for the \PH{A}{1g}(9) R mode. In Fig.~\ref{fig_5} (a,b) we show the frequencies and relative amplitudes of the R mode motions as a function of the drive frequency $\Omega$, obtained using the parameters for the \PH{A}{2u}(17) (IR) -- \PH{A}{1g}(9) (R) coupled phonon modes. Note that the only effect of the external driving amplitude, $E_\mathrm{drive}$, is to scale the amplitude of the motion. We see that for drive frequencies close to the \unit[17]{THz} eigenfrequency of the IR mode, the frequencies of the R mode components range from sub THz to \unit[40]{THz} (note the logarithmic scale in the lower part of Fig.~\ref{fig_5} (a)), with the highest frequency components at around 34 THz being twice the renormalized IR mode eigenfrequency (blue line), twice the driver frequency (green dashed line) and the sum of the renormalized IR mode and driver frequencies (red dashed-dotted line). The renormalized R mode frequency is close to \unit[10]{THz}, and like the renormalized IR mode frequency is independent of the drive frequency. The frequency of the lowest frequency component of the motion is given by the difference between the frequency of the driver and the IR mode eigenfrequency, and as a result it has a strong dependence on the drive frequency, becoming small as the drive frequency approaches the eigenfrequency of the IR mode (note that the divergence when the drive frequency equals the eigenfrequency of the IR mode is not physical, and arises because of the absence of damping in our simulations.) 

In Fig.~\ref{fig_5} (b) we show the relative amplitudes of each frequency component normalized to the time-independent displacement $\xi_{\mathrm{R0}}$ of $\xi_\mathrm{R}(t)$ which we set to \unit[100]{\%}. We see a large spread in amplitudes for the different components of the motion, with the motions with the renormalized IR and R eigenfrequencies having the largest amplitudes, in the order of 10 to \unit[20]{\%} of $\xi_{\mathrm{R0}}$, as well as minimal dependence on the drive frequency. The other three motion components, whose frequencies depend explicitly on the drive frequency, have strongly drive-frequency-dependent amplitudes, as expected. Of these, the high-frequency 2$\Omega$ motion has the smallest amplitude followed by the $\Omega+\noIR$ motion, with the slow $\Omega-\noIR$ motion having the largest amplitude, becoming similar in size to the $\noIR$ and $\noR$ motions in the vicinity of the eigenfrequency of IR mode. Again, the divergence when the mode frequency matches the driver frequency results from the absence of damping in our model, and so we do not analyze this point in detail. 

To summarize this section, we find that, in addition to the time-independent offset $\xi_{\mathrm{R0}}$ of the R mode induced by its quadratic-linear coupling to the IR mode,  the R mode has a complex oscillatory motion made up of different frequencies. The largest amplitude motions have high frequencies, set by the $\noIR$ and $\noR$ frequencies. Close to resonance between the drive and IR-mode frequencies, an additional component of the motion with a slower frequency $\Omega-\noIR$ also develops a significant amplitude. This slow motion is particularly interesting since it is tunable in amplitude and frequency by the external driver; in the next section we will explore how it can be exploited to engineer the spin dynamics. 

\section{Spin dynamics}
\label{SpinDynamics}
Next we discuss how the structural modulations we described above drive the spin dynamics, by combining our findings for the structural dynamics with those for the spin-phonon coupling. The time-dependent exchange modulation induced by the structural modulation is obtained by combining Eqn.~\eqref{eq_single_exchange} for $J_n(\xi)$ with Eqn.~\eqref{eq_Rdynamics} for $\xi(t)$ to yield the $J_n(\xi(t))$. \rev{We include only the modulations caused by the \PH{A}{1g}(9) R mode. While this mode is driven by the excitation of the \PH{A}{2u}(17) (IR) mode, the latter has negligible effect on the exchange interactions, and so the time-dependent magnetic exchange modulations are dominated by $J_n(\xi_\mathrm{R}(t))$}. 

\rev{For the spin-dynamics we consider a single Cr$_2$O$_3$ unit-cell with four magnetic Cr sites and periodic boundary conditions. Rewriting the Heisenberg Hamiltonian of Eqn.~\eqref{eq_cr2o3exchanges} we obtain} 
\begin{align}
	H^\mathrm{mag,exch}_\mathrm{fac}(\xi_\mathrm{R}(t))=&\: \tilde{J}_1(\xi_\mathrm{R}(t))(S_1 \cdot S_2+S_3 \cdot S_4)\nonumber \\
	&+\tilde{J}_2(\xi_\mathrm{R}(t))(S_1 \cdot S_4+S_2 \cdot S_3) \\
	&+\tilde{J}_3(\xi_\mathrm{R}(t))(S_1 \cdot S_3+S_2 \cdot S_4)\nonumber ,
\end{align}
\rev{where the net magnetic exchange interactions $\tilde{J}_i$ are given by}
\begin{align}
\tilde{J}_1(\xi_\mathrm{R}(t))=&J_1(\xi_\mathrm{R}(t))+3J_3(\xi_\mathrm{R}(t))\nonumber\\
\tilde{J}_2(\xi_\mathrm{R}(t))=&3J_2(\xi_\mathrm{R}(t))+J_5(\xi_\mathrm{R}(t))\label{eq_renorm_exhanges}\\
\tilde{J}_3(\xi_\mathrm{R}(t))=&6J_4(\xi_\mathrm{R}(t))\nonumber \; 
\end{align}
and $\vec{S}_1$ to $\vec{S}_4$ are the four classical spins in the unit cell as shown in Fig.~\ref{fig_1} (a).
Our \rev{full} magnetic Hamiltonian is then 
\begin{equation}
\begin{aligned}
\label{eq_LLD_0}
	\displaystyle {\mathcal H^{\mathrm{mag}}}(t) =& H^\mathrm{mag,exch}_\mathrm{fac}(\xi_\mathrm{R}(t)) +  D\displaystyle\sum_{i=1}^4 (S^{\mathrm{z}}_i)^2
\end{aligned}
\end{equation}
where \rev{$\xi_{\mathrm{R}}(t))$ denotes the specific time-dependent exchange interaction strength (see Eqn.~\eqref{eq_renorm_exhanges} and Table \ref{tab_3}) and $D$ is the MCA energy which we fixed to the computed equilibrium value.} 

\begin{figure}[tb]
\includegraphics[width=1\columnwidth]{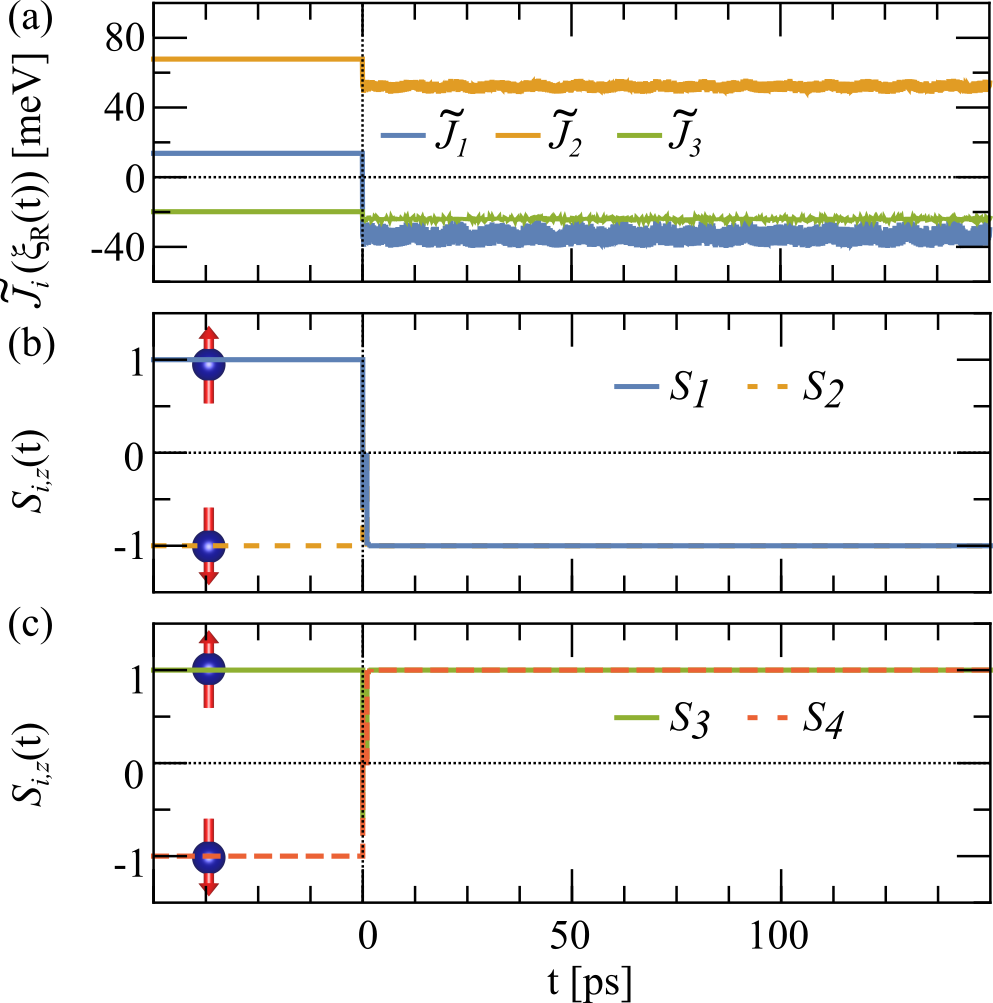}
\caption{\label{fig_6}\rev{Spin dynamics in the phonon-driven state of \CrO. (a) Modulation of the magnetic exchange interaction $\tilde{J}_i(\xi_{\mathrm{R}}(t)$) with $\xi_{\mathrm{R}}(t)$ derived from Eqn.~\eqref{eq_Rdynamics}. The laser field, $E(t)$, is switched on at time $t=0$, with $E_\mathrm{drive}=$\unit[0.6]{MV/cm} at $\Omega=$\unit[16.900]{THz}. The change in sign of the average exchange for $\tilde{J}_1$ is clearly visible; note that because of the fast oscillating components ($\omega\geq$\unit[30]{THz}) of the $\xi_\mathrm{R}$ motion the time-dependent exchange interaction can not be resolved on this scale. (b,c) Time dependence of the $z$-components (normalized to their ground-state values) of the four spin magnetic moments in the \CrO\ unit cell with the labeling corresponding to Fig.~\ref{fig_1} (a). The blue spheres (Cr atoms) and red arrows (spins) represent the  \CrO\ magnetic ground state.}}
\end{figure}

\rev{We next calculate the classical magnetization dynamics using the Landau-Lifshitz-Gilbert equation \cite{Landau:1935tr,Gilbert:1955ta,Gilbert:2004} within an atomistic approach\cite{ChubykaloFesenko:2006hq,Skubic:2008gs}}
\begin{equation}
\begin{aligned}
\label{eq_LLD_1}
\displaystyle \frac{d\vec{S}_i}{dt}=& - \frac{\gamma}{1+\alpha^2}\left[\vec{S}_i \times \vec{H}_i^{\mathrm{eff}}(t)\right] \\
& - \frac{\alpha \gamma}{1+\alpha^2}\left[\vec{S}_i \times \left[\vec{S}_i \times \vec{H}_i^{\mathrm{eff}}(t)\right]\right].
\end{aligned}
\end{equation}
Here $\vec{H}_i^{\mathrm{eff}}(t)=-\frac{1}{\mu_\mathrm{B}}\frac{\delta {\mathcal H_\mathrm{fac}^\mathrm{mag,exch}}(t)}{\delta \vec{S}_i} \;$ with $\mu_{\mathrm{B}}$ the Bohr magneton, $\gamma$ is the electron gyromagnetic ratio and $\alpha$ is the Gilbert damping. We take the value of Gilbert damping for \CrO, $\alpha\approx \frac{\gamma}{2\pi \nu}\Delta B_{\mathrm{pp}} \approx 0.07 $, estimated from spectral line widths measured at room temperature using electron paramagnetic resonance\cite{Stone:1971gj}. 

Next, we calculate $\tilde{J}_i(\xi_{\mathrm{R}}(t))$ when the structure is modified by the quadratic-linear coupling of the \PH{A}{1g}(9) and \PH{A}{2u}(17) phonon modes. We excite the latter using a continuous field of strength of \rev{$E_\mathrm{drive}=$\unit[0.6]{MV/cm}} oscillating at a frequency of \unit[16.9]{THz}; the result is shown in Fig.~\ref{fig_6} (a). Note that we include a noise corresponding to a temperature of \unit[0.1]{K} in our spin-dynamics simulation to prevent the system from becoming stuck in shallow metastable  minima\cite{Sukhov:2009gn}. 

Before the mode is excited (at $t=0$), all $\tilde{J}_i$ are constant, with $\tilde{J}_1$ and $\tilde{J}_2$ positive and $\tilde{J}_3$ negative. When the oscillating field is applied, the frequency-dependent induced structural changes described in the previous section change $\tilde{J}_i$ corresponding to the changes in bond lengths and angles. We see that, while the magnetic exchange interactions oscillate, the {\it average} magnetic exchange interactions $\tilde{J}_1$ change sign to negative, reflecting a net ferromagnetic interaction between the nearest neighbor sites. Since the next-nearest neighbor interaction $\tilde{J}_2$ still prefers an AFM alignment the system does not become fully \rev{FM but instead adopts the AFM$_1$ state with its $(\uparrow,\uparrow,\downarrow,\downarrow)$ ordering of magnetic moments on the Cr sites. This is consistent with the cross-over to the AFM$_1$ state } with increasing \PH{A}{1g}(9) amplitude that we saw in the first part of this paper. 
 
\begin{figure}[tb]
\includegraphics[width=1\columnwidth]{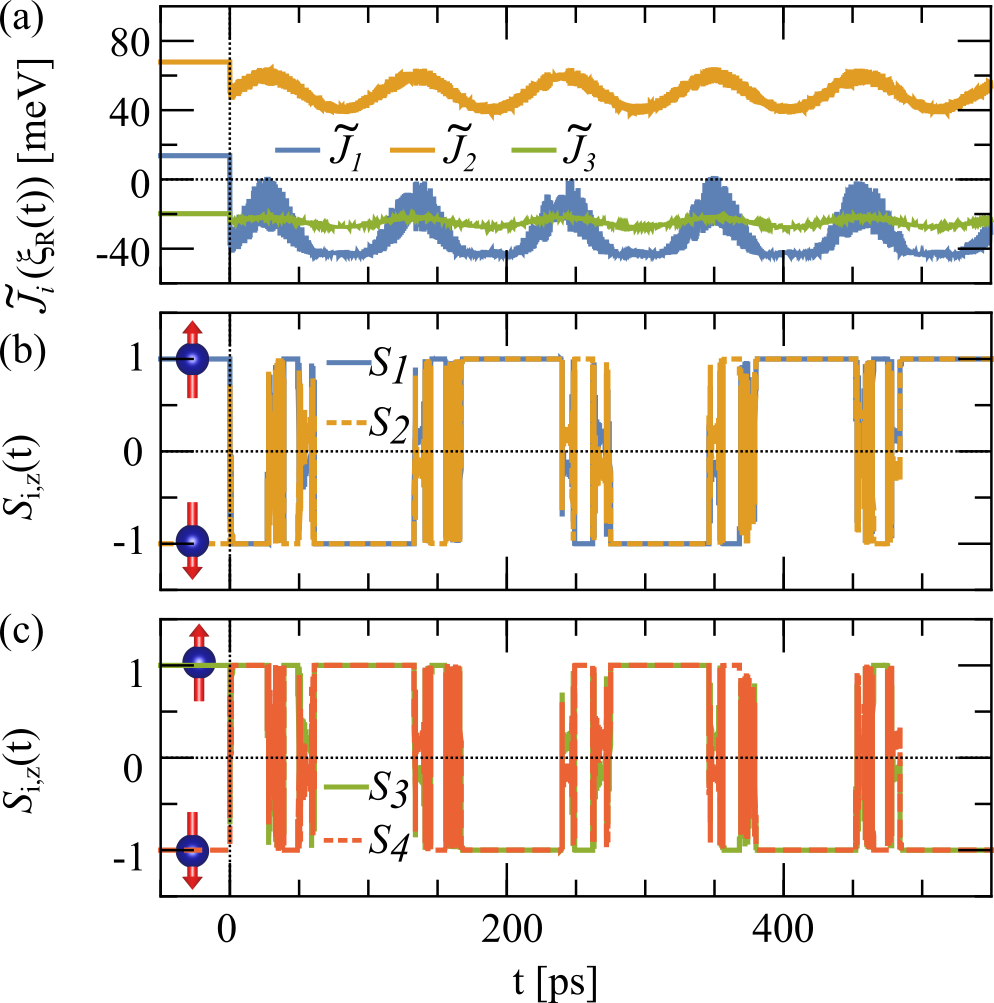}
\caption{\label{fig_7} \rev{Spin dynamics in the phonon-driven state of \CrO. (a) Modulation of the magnetic exchange interaction $\tilde{J}_i(\xi_{\mathrm{R}}(t))$ with $\xi_{\mathrm{R}}$ derived from Eqn.~\eqref{eq_Rdynamics}. The laser field $E(t)$ is switched on at time $t=0$, with $E_\mathrm{drive}=$\unit[0.6]{MV/cm} at $\Omega=$\unit[16.995]{THz}. Setting the excitation frequency closer to  resonance induces  a slow, large-amplitude modulation component in the $\xi_\mathrm{R}$ motion, which becomes resolvable in the time-dependent magnetic exchange. (b,c) Time dependence of the $z$-component of the four spin magnetic moments in the \CrO\ unit cell with the labeling corresponding to Fig.~\ref{fig_1} (a) and the spin magnitudes normalized to their static ground-state values. We illustrate the \CrO\ magnetic ground state by the blue spheres representing the Cr atoms with the arrows showing the magnetic moments.}}
\end{figure}

The remaining panels of Fig.~\ref{fig_6} show the response of the spin system to this modification of $\tilde{J}_i$, with (b) and (c) showing the time evolution of the $z$-component of magnetization of the individual Cr ions and (d) that of the total spin moment of the unit cell, $S_{tot,z}(t)=\frac{1}{N}\sum_{i=1}^N S_{i,\mathrm{z}}(t)$. The spins react to the change in their average exchange modulation and \rev{reorient on the same time scale as the J-oscillations into the new AFM arrangement}. Note that this process is at the same speed of the period of the exchange excitation \rev{oscillations}, $2\pi/\noR\approx$\unit[0.1] {ps}.  The AFM arrangement then achieves a steady state without further dynamical evolution provided that the displacement of the \PH{A}{1g}(9) continues by excitation of the \PH{A}{2u}(17) phonon mode. 

Next, we exploit our finding from Section~\ref{NLLD} that a component of the R mode motion can be tuned to low frequency with an increased amplitude by selecting a drive frequency $\Omega$ close to resonance. \rev{In Fig.~\ref{fig_7} (a) we show the time dependence of $\tilde{J}_i$, calculated for the combination of \PH{A}{1g}(9) and \PH{A}{2u}(17) modes using the analytical solution of Eqn.~\eqref{eq_Rdynamics}, this time with the driving frequency, $\Omega=$\unit[16.995]{THz}, close to resonance.  It is clear that the oscillation frequency of the exchange interactions develops a significant slow component with a frequency  around $10$~GHz. The resulting spin dynamics are depicted   in the lower panels of Fig.~\ref{fig_7}. In contrast to the case shown in Fig.~\ref{fig_6}, a steady AFM$_1$ state is not achieved on pumping, and instead the spins exhibit a flipping between up and down alignment.} Again the spin dynamics behavior persists as long as the phonon mode is driven. 

In conclusion, our calculations indicate that the quadratic-linear coupling between the \PH{A}{1g}(9) and \PH{A}{2u}(17) modes leads to a reversal of the average value of the nearest-neighbor exchange between the Cr ions when the optical \PH{A}{2u}(17) mode is continuously excited with sufficiently large amplitude. Depending on the closeness of the excitation laser frequency to the eigenfrequency of the \PH{A}{2u}(17) mode, the additional oscillatory component of $\tilde{J}_i(t)$ can be either fast or slow. In the first case the system responds with a steady-state change in its magnetization to a ferromagnetic state; in the second an alternating switching occurs on a tens of picoseconds time scale. We note that the two limits shown here represent a small fraction of spin-dynamic possibilities, with the tuning of the drive frequency relative to the resonance, as well as on-off schemes for the excitation, offering the potential to modulate the exchange interactions in multiple complex ways.  

\section{Summary}

We  calculated the structural and magnetic responses of chromium oxide, \CrO, to intense excitation of its optically active phonon modes. Using a general spin-lattice Hamiltonian, with parameters calculated from first principles, we showed that the quasi-static structural distortion introduced through the non-linear phonon-phonon interaction can change the magnetic state from its equilibrium antiferromagnetic to a \rev{new antiferromagnetic ordering with ferromagnetically coupled nearest-neighbor spins.} This transition is driven by the change in nearest-neighbor magnetic exchange interaction when the Cr-Cr separation is modified through non-linear coupling of the optical phonons to a symmetry-conserving \PH{A}{1g} Raman-active mode.\rev{ The new antiferromagnetic ground state persists for as long as the system is continuously excited, provided that the excitation frequency is faster than the magnetic relaxation time. }

Regarding dynamics, we find that the motion of the excited optical modes and coupled Raman-active mode can be decomposed into several different frequencies which depend strongly on the difference between the excitation and resonance frequencies. This sensitivity of the response to the input frequency allows selection of complex vibrational frequency patterns which can lead to additional components in the spin dynamics, for example flips of the Cr spin lattice. 

We emphasize that we explored in this work a minimal model of phonon-driven spin dynamics, and we expect that extensions of the model will reveal yet more complex physics, such as dynamically frustrated or spin-spiral states. We hope that our work will inspire additional theoretical and experimental studies to uncover the rich behavior of coupled magneto-phononic systems. 

\section{Acknowledgments}
This work was supported financially by ETH Zurich, the ERC Advanced Grant program, No. 291151 (MF, CK and NAS), the ERC under the European Union's Seventh Framework Programme (FP7/2007-2013) / ERC Grant Agreement n$^\circ$ 319286 (Q-MAC) and by the DFG through SFB762 and TRR227. Calculations were performed at the Swiss National Supercomputing Centre (CSCS) under project ID s624. MF and NAS thank Andrea Cavalleri and Tobia F. Nova for useful discussions.

\bibliography{fullbib}

\begin{thebibliography}{52}
\expandafter\ifx\csname natexlab\endcsname\relax\def\natexlab#1{#1}\fi
\expandafter\ifx\csname bibnamefont\endcsname\relax
  \def\bibnamefont#1{#1}\fi
\expandafter\ifx\csname bibfnamefont\endcsname\relax
  \def\bibfnamefont#1{#1}\fi
\expandafter\ifx\csname citenamefont\endcsname\relax
  \def\citenamefont#1{#1}\fi
\expandafter\ifx\csname url\endcsname\relax
  \def\url#1{\texttt{#1}}\fi
\expandafter\ifx\csname urlprefix\endcsname\relax\def\urlprefix{URL }\fi
\providecommand{\bibinfo}[2]{#2}
\providecommand{\eprint}[2][]{\url{#2}}

\bibitem[{\citenamefont{F{\"o}rst
  et~al.}(2011{\natexlab{a}})\citenamefont{F{\"o}rst, Manzoni, Kaiser, Tomioka,
  Tokura, Merlin, and Cavalleri}}]{Forst:2011ep}
\bibinfo{author}{\bibfnamefont{M.}~\bibnamefont{F{\"o}rst}},
  \bibinfo{author}{\bibfnamefont{C.}~\bibnamefont{Manzoni}},
  \bibinfo{author}{\bibfnamefont{S.}~\bibnamefont{Kaiser}},
  \bibinfo{author}{\bibfnamefont{Y.}~\bibnamefont{Tomioka}},
  \bibinfo{author}{\bibfnamefont{Y.}~\bibnamefont{Tokura}},
  \bibinfo{author}{\bibfnamefont{R.}~\bibnamefont{Merlin}}, \bibnamefont{and}
  \bibinfo{author}{\bibfnamefont{A.}~\bibnamefont{Cavalleri}},
  \bibinfo{journal}{Nature Phys.} \textbf{\bibinfo{volume}{7}},
  \bibinfo{pages}{854} (\bibinfo{year}{2011}{\natexlab{a}}).

\bibitem[{\citenamefont{Subedi et~al.}(2014)\citenamefont{Subedi, Cavalleri,
  and Georges}}]{Subedi:2014ik}
\bibinfo{author}{\bibfnamefont{A.}~\bibnamefont{Subedi}},
  \bibinfo{author}{\bibfnamefont{A.}~\bibnamefont{Cavalleri}},
  \bibnamefont{and} \bibinfo{author}{\bibfnamefont{A.}~\bibnamefont{Georges}},
  \bibinfo{journal}{Phys. Rev. B} \textbf{\bibinfo{volume}{89}},
  \bibinfo{pages}{220301} (\bibinfo{year}{2014}).

\bibitem[{\citenamefont{Rini et~al.}(2007)\citenamefont{Rini, Tobey, Dean,
  Itatani, Tomioka, Tokura, Schoenlein, and Cavalleri}}]{Rini:2007hca}
\bibinfo{author}{\bibfnamefont{M.}~\bibnamefont{Rini}},
  \bibinfo{author}{\bibfnamefont{R.}~\bibnamefont{Tobey}},
  \bibinfo{author}{\bibfnamefont{N.}~\bibnamefont{Dean}},
  \bibinfo{author}{\bibfnamefont{J.}~\bibnamefont{Itatani}},
  \bibinfo{author}{\bibfnamefont{Y.}~\bibnamefont{Tomioka}},
  \bibinfo{author}{\bibfnamefont{Y.}~\bibnamefont{Tokura}},
  \bibinfo{author}{\bibfnamefont{R.~W.} \bibnamefont{Schoenlein}},
  \bibnamefont{and}
  \bibinfo{author}{\bibfnamefont{A.}~\bibnamefont{Cavalleri}},
  \bibinfo{journal}{Nature} \textbf{\bibinfo{volume}{449}}, \bibinfo{pages}{72}
  (\bibinfo{year}{2007}).

\bibitem[{\citenamefont{Caviglia et~al.}(2012)\citenamefont{Caviglia,
  Scherwitzl, Popovich, Hu, Bromberger, Singla, Mitrano, Hoffmann, Kaiser,
  Zubko et~al.}}]{Caviglia:2012eaa}
\bibinfo{author}{\bibfnamefont{A.~D.} \bibnamefont{Caviglia}},
  \bibinfo{author}{\bibfnamefont{R.}~\bibnamefont{Scherwitzl}},
  \bibinfo{author}{\bibfnamefont{P.}~\bibnamefont{Popovich}},
  \bibinfo{author}{\bibfnamefont{W.}~\bibnamefont{Hu}},
  \bibinfo{author}{\bibfnamefont{H.}~\bibnamefont{Bromberger}},
  \bibinfo{author}{\bibfnamefont{R.}~\bibnamefont{Singla}},
  \bibinfo{author}{\bibfnamefont{M.}~\bibnamefont{Mitrano}},
  \bibinfo{author}{\bibfnamefont{M.~C.} \bibnamefont{Hoffmann}},
  \bibinfo{author}{\bibfnamefont{S.}~\bibnamefont{Kaiser}},
  \bibinfo{author}{\bibfnamefont{P.}~\bibnamefont{Zubko}},
  \bibnamefont{et~al.}, \bibinfo{journal}{Phys. Rev. Lett.}
  \textbf{\bibinfo{volume}{108}}, \bibinfo{pages}{136801}
  (\bibinfo{year}{2012}).

\bibitem[{\citenamefont{Esposito et~al.}(2017)\citenamefont{Esposito, Fechner,
  Mankowsky, Lemke, M, Glownia, Nakamura, Kawasaki, Tokura, Staub
  et~al.}}]{Esposito:2017wy}
\bibinfo{author}{\bibfnamefont{V.}~\bibnamefont{Esposito}},
  \bibinfo{author}{\bibfnamefont{M.}~\bibnamefont{Fechner}},
  \bibinfo{author}{\bibfnamefont{R.}~\bibnamefont{Mankowsky}},
  \bibinfo{author}{\bibfnamefont{H.}~\bibnamefont{Lemke}},
  \bibinfo{author}{\bibfnamefont{C.}~\bibnamefont{M}},
  \bibinfo{author}{\bibfnamefont{J.~M.} \bibnamefont{Glownia}},
  \bibinfo{author}{\bibfnamefont{M.}~\bibnamefont{Nakamura}},
  \bibinfo{author}{\bibfnamefont{M.}~\bibnamefont{Kawasaki}},
  \bibinfo{author}{\bibfnamefont{Y.}~\bibnamefont{Tokura}},
  \bibinfo{author}{\bibfnamefont{U.}~\bibnamefont{Staub}},
  \bibnamefont{et~al.}, \bibinfo{journal}{Phys. Rev. Lett.}
  \textbf{\bibinfo{volume}{118}}, \bibinfo{pages}{247601}
  (\bibinfo{year}{2017}).

\bibitem[{\citenamefont{Kaiser et~al.}(2014)\citenamefont{Kaiser, Hunt,
  Nicoletti, Hu, Gierz, Liu, Le~Tacon, Loew, Haug, Keimer
  et~al.}}]{Kaiser:2014de}
\bibinfo{author}{\bibfnamefont{S.}~\bibnamefont{Kaiser}},
  \bibinfo{author}{\bibfnamefont{C.~R.} \bibnamefont{Hunt}},
  \bibinfo{author}{\bibfnamefont{D.}~\bibnamefont{Nicoletti}},
  \bibinfo{author}{\bibfnamefont{W.}~\bibnamefont{Hu}},
  \bibinfo{author}{\bibfnamefont{I.}~\bibnamefont{Gierz}},
  \bibinfo{author}{\bibfnamefont{H.~Y.} \bibnamefont{Liu}},
  \bibinfo{author}{\bibfnamefont{M.}~\bibnamefont{Le~Tacon}},
  \bibinfo{author}{\bibfnamefont{T.}~\bibnamefont{Loew}},
  \bibinfo{author}{\bibfnamefont{D.}~\bibnamefont{Haug}},
  \bibinfo{author}{\bibfnamefont{B.}~\bibnamefont{Keimer}},
  \bibnamefont{et~al.}, \bibinfo{journal}{Phys. Rev. B}
  \textbf{\bibinfo{volume}{89}}, \bibinfo{pages}{184516}
  (\bibinfo{year}{2014}).

\bibitem[{\citenamefont{Hu et~al.}(2014)\citenamefont{Hu, Kaiser, Nicoletti,
  Hunt, Gierz, Hoffmann, Le~Tacon, Loew, Keimer, and Cavalleri}}]{Hu:2014cg}
\bibinfo{author}{\bibfnamefont{W.}~\bibnamefont{Hu}},
  \bibinfo{author}{\bibfnamefont{S.}~\bibnamefont{Kaiser}},
  \bibinfo{author}{\bibfnamefont{D.}~\bibnamefont{Nicoletti}},
  \bibinfo{author}{\bibfnamefont{C.~R.} \bibnamefont{Hunt}},
  \bibinfo{author}{\bibfnamefont{I.}~\bibnamefont{Gierz}},
  \bibinfo{author}{\bibfnamefont{M.~C.} \bibnamefont{Hoffmann}},
  \bibinfo{author}{\bibfnamefont{M.}~\bibnamefont{Le~Tacon}},
  \bibinfo{author}{\bibfnamefont{T.}~\bibnamefont{Loew}},
  \bibinfo{author}{\bibfnamefont{B.}~\bibnamefont{Keimer}}, \bibnamefont{and}
  \bibinfo{author}{\bibfnamefont{A.}~\bibnamefont{Cavalleri}},
  \bibinfo{journal}{Nature Mater.} \textbf{\bibinfo{volume}{13}},
  \bibinfo{pages}{705} (\bibinfo{year}{2014}).

\bibitem[{\citenamefont{Mitrano et~al.}(2016)\citenamefont{Mitrano, Cantaluppi,
  Nicoletti, Kaiser, Perucchi, Lupi, Di~Pietro, Pontiroli, Ricc{\`o}, Clark
  et~al.}}]{Mitrano:2016fr}
\bibinfo{author}{\bibfnamefont{M.}~\bibnamefont{Mitrano}},
  \bibinfo{author}{\bibfnamefont{A.}~\bibnamefont{Cantaluppi}},
  \bibinfo{author}{\bibfnamefont{D.}~\bibnamefont{Nicoletti}},
  \bibinfo{author}{\bibfnamefont{S.}~\bibnamefont{Kaiser}},
  \bibinfo{author}{\bibfnamefont{A.}~\bibnamefont{Perucchi}},
  \bibinfo{author}{\bibfnamefont{S.}~\bibnamefont{Lupi}},
  \bibinfo{author}{\bibfnamefont{P.}~\bibnamefont{Di~Pietro}},
  \bibinfo{author}{\bibfnamefont{D.}~\bibnamefont{Pontiroli}},
  \bibinfo{author}{\bibfnamefont{M.}~\bibnamefont{Ricc{\`o}}},
  \bibinfo{author}{\bibfnamefont{S.~R.} \bibnamefont{Clark}},
  \bibnamefont{et~al.}, \textbf{\bibinfo{volume}{530}}, \bibinfo{pages}{461}
  (\bibinfo{year}{2016}).

\bibitem[{\citenamefont{Mankowsky et~al.}(2014)\citenamefont{Mankowsky, Subedi,
  F{\"o}rst, Mariager, M, Lemke, Robinson, Glownia, Minitti, Frano
  et~al.}}]{Mankowsky:2014vt}
\bibinfo{author}{\bibfnamefont{R.}~\bibnamefont{Mankowsky}},
  \bibinfo{author}{\bibfnamefont{A.}~\bibnamefont{Subedi}},
  \bibinfo{author}{\bibfnamefont{M.}~\bibnamefont{F{\"o}rst}},
  \bibinfo{author}{\bibfnamefont{S.~O.} \bibnamefont{Mariager}},
  \bibinfo{author}{\bibfnamefont{C.}~\bibnamefont{M}},
  \bibinfo{author}{\bibfnamefont{H.~T.} \bibnamefont{Lemke}},
  \bibinfo{author}{\bibfnamefont{J.~S.} \bibnamefont{Robinson}},
  \bibinfo{author}{\bibfnamefont{J.~M.} \bibnamefont{Glownia}},
  \bibinfo{author}{\bibfnamefont{M.~P.} \bibnamefont{Minitti}},
  \bibinfo{author}{\bibfnamefont{A.}~\bibnamefont{Frano}},
  \bibnamefont{et~al.}, \bibinfo{journal}{Nature}
  \textbf{\bibinfo{volume}{516}}, \bibinfo{pages}{71} (\bibinfo{year}{2014}).

\bibitem[{\citenamefont{Fechner and Spaldin}(2016)}]{Fechner:2016tu}
\bibinfo{author}{\bibfnamefont{M.}~\bibnamefont{Fechner}} \bibnamefont{and}
  \bibinfo{author}{\bibfnamefont{N.~A.} \bibnamefont{Spaldin}},
  \bibinfo{journal}{Phys. Rev. B} \textbf{\bibinfo{volume}{94}},
  \bibinfo{pages}{134307} (\bibinfo{year}{2016}).

\bibitem[{\citenamefont{Mankowsky
  et~al.}(2017{\natexlab{a}})\citenamefont{Mankowsky, Fechner, F{\"o}rst, von
  Hoegen, Porras, Loew, Dakovski, Seaberg, M{\"o}ller, Coslovich
  et~al.}}]{Mankowsky:2017br}
\bibinfo{author}{\bibfnamefont{R.}~\bibnamefont{Mankowsky}},
  \bibinfo{author}{\bibfnamefont{M.}~\bibnamefont{Fechner}},
  \bibinfo{author}{\bibfnamefont{M.}~\bibnamefont{F{\"o}rst}},
  \bibinfo{author}{\bibfnamefont{A.}~\bibnamefont{von Hoegen}},
  \bibinfo{author}{\bibfnamefont{J.}~\bibnamefont{Porras}},
  \bibinfo{author}{\bibfnamefont{T.}~\bibnamefont{Loew}},
  \bibinfo{author}{\bibfnamefont{G.~L.} \bibnamefont{Dakovski}},
  \bibinfo{author}{\bibfnamefont{M.}~\bibnamefont{Seaberg}},
  \bibinfo{author}{\bibfnamefont{S.}~\bibnamefont{M{\"o}ller}},
  \bibinfo{author}{\bibfnamefont{G.}~\bibnamefont{Coslovich}},
  \bibnamefont{et~al.}, \bibinfo{journal}{Structural Dynamics}
  \textbf{\bibinfo{volume}{4}}, \bibinfo{pages}{044007}
  (\bibinfo{year}{2017}{\natexlab{a}}).

\bibitem[{\citenamefont{Gu and Rondinelli}(2017)}]{Gu:2017dm}
\bibinfo{author}{\bibfnamefont{M.}~\bibnamefont{Gu}} \bibnamefont{and}
  \bibinfo{author}{\bibfnamefont{J.~M.} \bibnamefont{Rondinelli}},
  \bibinfo{journal}{Phys. Rev. B} \textbf{\bibinfo{volume}{95}},
  \bibinfo{pages}{024109} (\bibinfo{year}{2017}).

\bibitem[{\citenamefont{Juraschek
  et~al.}(2017{\natexlab{a}})\citenamefont{Juraschek, Fechner, and
  Spaldin}}]{Juraschek:2017ic}
\bibinfo{author}{\bibfnamefont{D.~M.} \bibnamefont{Juraschek}},
  \bibinfo{author}{\bibfnamefont{M.}~\bibnamefont{Fechner}}, \bibnamefont{and}
  \bibinfo{author}{\bibfnamefont{N.~A.} \bibnamefont{Spaldin}},
  \bibinfo{journal}{Phys. Rev. Lett.} \textbf{\bibinfo{volume}{118}},
  \bibinfo{pages}{054101} (\bibinfo{year}{2017}{\natexlab{a}}).

\bibitem[{\citenamefont{Subedi}(2015)}]{Subedi:2015dw}
\bibinfo{author}{\bibfnamefont{A.}~\bibnamefont{Subedi}},
  \bibinfo{journal}{Phys. Rev. B} \textbf{\bibinfo{volume}{92}},
  \bibinfo{pages}{214303} (\bibinfo{year}{2015}).

\bibitem[{\citenamefont{Juraschek
  et~al.}(2017{\natexlab{b}})\citenamefont{Juraschek, Fechner, Balatsky, and
  Spaldin}}]{Juraschek:2017ur}
\bibinfo{author}{\bibfnamefont{D.~M.} \bibnamefont{Juraschek}},
  \bibinfo{author}{\bibfnamefont{M.}~\bibnamefont{Fechner}},
  \bibinfo{author}{\bibfnamefont{A.~V.} \bibnamefont{Balatsky}},
  \bibnamefont{and} \bibinfo{author}{\bibfnamefont{N.~A.}
  \bibnamefont{Spaldin}}, \bibinfo{journal}{Phys. Rev. Mat.}
  \textbf{\bibinfo{volume}{1}}, \bibinfo{pages}{014401}
  (\bibinfo{year}{2017}{\natexlab{b}}).

\bibitem[{\citenamefont{Mankowsky
  et~al.}(2017{\natexlab{b}})\citenamefont{Mankowsky, von Hoegen, F{\"o}rst,
  and Cavalleri}}]{Mankowsky:2017ur}
\bibinfo{author}{\bibfnamefont{R.}~\bibnamefont{Mankowsky}},
  \bibinfo{author}{\bibfnamefont{A.}~\bibnamefont{von Hoegen}},
  \bibinfo{author}{\bibfnamefont{M.}~\bibnamefont{F{\"o}rst}},
  \bibnamefont{and}
  \bibinfo{author}{\bibfnamefont{A.}~\bibnamefont{Cavalleri}},
  \bibinfo{journal}{Phys. Rev. Lett.} \textbf{\bibinfo{volume}{118}},
  \bibinfo{pages}{197601} (\bibinfo{year}{2017}{\natexlab{b}}).

\bibitem[{\citenamefont{F{\"o}rst
  et~al.}(2011{\natexlab{b}})\citenamefont{F{\"o}rst, Tobey, Wall, Bromberger,
  Khanna, Cavalieri, Chuang, Lee, Moore, Schlotter et~al.}}]{Forst:2011kl}
\bibinfo{author}{\bibfnamefont{M.}~\bibnamefont{F{\"o}rst}},
  \bibinfo{author}{\bibfnamefont{R.~I.} \bibnamefont{Tobey}},
  \bibinfo{author}{\bibfnamefont{S.}~\bibnamefont{Wall}},
  \bibinfo{author}{\bibfnamefont{H.}~\bibnamefont{Bromberger}},
  \bibinfo{author}{\bibfnamefont{V.}~\bibnamefont{Khanna}},
  \bibinfo{author}{\bibfnamefont{A.~L.} \bibnamefont{Cavalieri}},
  \bibinfo{author}{\bibfnamefont{Y.~D.} \bibnamefont{Chuang}},
  \bibinfo{author}{\bibfnamefont{W.~S.} \bibnamefont{Lee}},
  \bibinfo{author}{\bibfnamefont{R.}~\bibnamefont{Moore}},
  \bibinfo{author}{\bibfnamefont{W.~F.} \bibnamefont{Schlotter}},
  \bibnamefont{et~al.}, \bibinfo{journal}{Phys. Rev. B}
  \textbf{\bibinfo{volume}{84}}, \bibinfo{pages}{241104}
  (\bibinfo{year}{2011}{\natexlab{b}}).

\bibitem[{\citenamefont{F{\"o}rst et~al.}(2015)\citenamefont{F{\"o}rst,
  Caviglia, Scherwitzl, Mankowsky, Zubko, Khanna, Bromberger, Wilkins, Chuang,
  Lee et~al.}}]{Forst:2015fv}
\bibinfo{author}{\bibfnamefont{M.}~\bibnamefont{F{\"o}rst}},
  \bibinfo{author}{\bibfnamefont{A.~D.} \bibnamefont{Caviglia}},
  \bibinfo{author}{\bibfnamefont{R.}~\bibnamefont{Scherwitzl}},
  \bibinfo{author}{\bibfnamefont{R.}~\bibnamefont{Mankowsky}},
  \bibinfo{author}{\bibfnamefont{P.}~\bibnamefont{Zubko}},
  \bibinfo{author}{\bibfnamefont{V.}~\bibnamefont{Khanna}},
  \bibinfo{author}{\bibfnamefont{H.}~\bibnamefont{Bromberger}},
  \bibinfo{author}{\bibfnamefont{S.~B.} \bibnamefont{Wilkins}},
  \bibinfo{author}{\bibfnamefont{Y.~D.} \bibnamefont{Chuang}},
  \bibinfo{author}{\bibfnamefont{W.~S.} \bibnamefont{Lee}},
  \bibnamefont{et~al.}, \bibinfo{journal}{Nature Mater.}
  \textbf{\bibinfo{volume}{14}}, \bibinfo{pages}{883} (\bibinfo{year}{2015}).

\bibitem[{\citenamefont{Nova et~al.}(2016)\citenamefont{Nova, Cartella,
  Cantaluppi, F{\"o}rst, Bossini, Mikhaylovskiy, Kimel, Merlin, and
  Cavalleri}}]{Nova:2016ja}
\bibinfo{author}{\bibfnamefont{T.~F.} \bibnamefont{Nova}},
  \bibinfo{author}{\bibfnamefont{A.}~\bibnamefont{Cartella}},
  \bibinfo{author}{\bibfnamefont{A.}~\bibnamefont{Cantaluppi}},
  \bibinfo{author}{\bibfnamefont{M.}~\bibnamefont{F{\"o}rst}},
  \bibinfo{author}{\bibfnamefont{D.}~\bibnamefont{Bossini}},
  \bibinfo{author}{\bibfnamefont{R.~V.} \bibnamefont{Mikhaylovskiy}},
  \bibinfo{author}{\bibfnamefont{A.~V.} \bibnamefont{Kimel}},
  \bibinfo{author}{\bibfnamefont{R.}~\bibnamefont{Merlin}}, \bibnamefont{and}
  \bibinfo{author}{\bibfnamefont{A.}~\bibnamefont{Cavalleri}},
  \bibinfo{journal}{Nature Phys.} \textbf{\bibinfo{volume}{13}},
  \bibinfo{pages}{132} (\bibinfo{year}{2016}).

\bibitem[{\citenamefont{Beaurepaire et~al.}(1996)\citenamefont{Beaurepaire,
  Merle, Daunois, and Bigot}}]{Beaurepaire:1996es}
\bibinfo{author}{\bibfnamefont{E.}~\bibnamefont{Beaurepaire}},
  \bibinfo{author}{\bibfnamefont{J.~C.} \bibnamefont{Merle}},
  \bibinfo{author}{\bibfnamefont{A.}~\bibnamefont{Daunois}}, \bibnamefont{and}
  \bibinfo{author}{\bibfnamefont{J.~Y.} \bibnamefont{Bigot}},
  \bibinfo{journal}{Phys. Rev. Lett.} \textbf{\bibinfo{volume}{76}},
  \bibinfo{pages}{4250} (\bibinfo{year}{1996}).

\bibitem[{\citenamefont{Kimel et~al.}(2002)\citenamefont{Kimel, Pisarev,
  Hohlfeld, and Rasing}}]{Kimel:2002ej}
\bibinfo{author}{\bibfnamefont{A.~V.} \bibnamefont{Kimel}},
  \bibinfo{author}{\bibfnamefont{R.~V.} \bibnamefont{Pisarev}},
  \bibinfo{author}{\bibfnamefont{J.}~\bibnamefont{Hohlfeld}}, \bibnamefont{and}
  \bibinfo{author}{\bibfnamefont{T.}~\bibnamefont{Rasing}},
  \bibinfo{journal}{Phys. Rev. Lett.} \textbf{\bibinfo{volume}{89}},
  \bibinfo{pages}{287401} (\bibinfo{year}{2002}).

\bibitem[{\citenamefont{Stanciu et~al.}(2007)\citenamefont{Stanciu, Hansteen,
  Kimel, Kirilyuk, Tsukamoto, Itoh, and Rasing}}]{Stanciu:2007fy}
\bibinfo{author}{\bibfnamefont{C.~D.} \bibnamefont{Stanciu}},
  \bibinfo{author}{\bibfnamefont{F.}~\bibnamefont{Hansteen}},
  \bibinfo{author}{\bibfnamefont{A.~V.} \bibnamefont{Kimel}},
  \bibinfo{author}{\bibfnamefont{A.}~\bibnamefont{Kirilyuk}},
  \bibinfo{author}{\bibfnamefont{A.}~\bibnamefont{Tsukamoto}},
  \bibinfo{author}{\bibfnamefont{A.}~\bibnamefont{Itoh}}, \bibnamefont{and}
  \bibinfo{author}{\bibfnamefont{T.}~\bibnamefont{Rasing}},
  \bibinfo{journal}{Phys. Rev. Lett.} \textbf{\bibinfo{volume}{99}},
  \bibinfo{pages}{047601} (\bibinfo{year}{2007}).

\bibitem[{\citenamefont{Kubacka et~al.}(2014)\citenamefont{Kubacka, Johnson,
  Hoffmann, Vicario, de~Jong, Beaud, Gr{\"u}bel, Huang, Huber, Patthey
  et~al.}}]{Kubacka:2014bf}
\bibinfo{author}{\bibfnamefont{T.}~\bibnamefont{Kubacka}},
  \bibinfo{author}{\bibfnamefont{J.~A.} \bibnamefont{Johnson}},
  \bibinfo{author}{\bibfnamefont{M.~C.} \bibnamefont{Hoffmann}},
  \bibinfo{author}{\bibfnamefont{C.}~\bibnamefont{Vicario}},
  \bibinfo{author}{\bibfnamefont{S.}~\bibnamefont{de~Jong}},
  \bibinfo{author}{\bibfnamefont{P.}~\bibnamefont{Beaud}},
  \bibinfo{author}{\bibfnamefont{S.}~\bibnamefont{Gr{\"u}bel}},
  \bibinfo{author}{\bibfnamefont{S.-W.} \bibnamefont{Huang}},
  \bibinfo{author}{\bibfnamefont{L.}~\bibnamefont{Huber}},
  \bibinfo{author}{\bibfnamefont{L.}~\bibnamefont{Patthey}},
  \bibnamefont{et~al.}, \bibinfo{journal}{Science}
  \textbf{\bibinfo{volume}{343}}, \bibinfo{pages}{1333} (\bibinfo{year}{2014}).

\bibitem[{\citenamefont{Landau and Lifshitz}(1935)}]{Landau:1935tr}
\bibinfo{author}{\bibfnamefont{L.}~\bibnamefont{Landau}} \bibnamefont{and}
  \bibinfo{author}{\bibfnamefont{E.}~\bibnamefont{Lifshitz}},
  \bibinfo{journal}{Phyz. Zeitsch. der Sow.} \textbf{\bibinfo{volume}{8}},
  \bibinfo{pages}{153} (\bibinfo{year}{1935}).

\bibitem[{\citenamefont{Gilbert and Kelly}(1955)}]{Gilbert:1955ta}
\bibinfo{author}{\bibfnamefont{T.~L.} \bibnamefont{Gilbert}} \bibnamefont{and}
  \bibinfo{author}{\bibfnamefont{J.~M.} \bibnamefont{Kelly}},
  \bibinfo{journal}{Conf. Magnetism and Magnetic Materials}
  \textbf{\bibinfo{volume}{14}}, \bibinfo{pages}{253} (\bibinfo{year}{1955}).

\bibitem[{\citenamefont{Wang et~al.}(2008)\citenamefont{Wang, Guo, and
  He}}]{Wang:2008dfa}
\bibinfo{author}{\bibfnamefont{C.}~\bibnamefont{Wang}},
  \bibinfo{author}{\bibfnamefont{G.-C.} \bibnamefont{Guo}}, \bibnamefont{and}
  \bibinfo{author}{\bibfnamefont{L.}~\bibnamefont{He}}, \bibinfo{journal}{Phys.
  Rev. B} \textbf{\bibinfo{volume}{77}}, \bibinfo{pages}{134113}
  (\bibinfo{year}{2008}).

\bibitem[{\citenamefont{Shi et~al.}(2009)\citenamefont{Shi, Wysocki, and
  Belashchenko}}]{Shi:2009jka}
\bibinfo{author}{\bibfnamefont{S.}~\bibnamefont{Shi}},
  \bibinfo{author}{\bibfnamefont{A.}~\bibnamefont{Wysocki}}, \bibnamefont{and}
  \bibinfo{author}{\bibfnamefont{K.~D.} \bibnamefont{Belashchenko}},
  \bibinfo{journal}{Phys. Rev. B} \textbf{\bibinfo{volume}{79}},
  \bibinfo{pages}{104404} (\bibinfo{year}{2009}).

\bibitem[{\citenamefont{Mostovoy et~al.}(2010)\citenamefont{Mostovoy,
  Scaramucci, Spaldin, and Delaney}}]{Mostovoy:2010ia}
\bibinfo{author}{\bibfnamefont{M.}~\bibnamefont{Mostovoy}},
  \bibinfo{author}{\bibfnamefont{A.}~\bibnamefont{Scaramucci}},
  \bibinfo{author}{\bibfnamefont{N.~A.} \bibnamefont{Spaldin}},
  \bibnamefont{and} \bibinfo{author}{\bibfnamefont{K.~T.}
  \bibnamefont{Delaney}}, \bibinfo{journal}{Phys. Rev. Lett.}
  \textbf{\bibinfo{volume}{105}}, \bibinfo{pages}{087202}
  (\bibinfo{year}{2010}).

\bibitem[{\citenamefont{Mu et~al.}(2014)\citenamefont{Mu, Wysocki, and
  Belashchenko}}]{Mu:2014gp}
\bibinfo{author}{\bibfnamefont{S.}~\bibnamefont{Mu}},
  \bibinfo{author}{\bibfnamefont{A.~L.} \bibnamefont{Wysocki}},
  \bibnamefont{and} \bibinfo{author}{\bibfnamefont{K.~D.}
  \bibnamefont{Belashchenko}}, \bibinfo{journal}{Phys. Rev. B}
  \textbf{\bibinfo{volume}{89}}, \bibinfo{pages}{174413}
  (\bibinfo{year}{2014}).

\bibitem[{\citenamefont{Kresse and Furthm{\"u}ller}(1996)}]{Kresse:1996vf}
\bibinfo{author}{\bibfnamefont{G.}~\bibnamefont{Kresse}} \bibnamefont{and}
  \bibinfo{author}{\bibfnamefont{J.}~\bibnamefont{Furthm{\"u}ller}},
  \bibinfo{journal}{Phys. Rev. B} \textbf{\bibinfo{volume}{54}},
  \bibinfo{pages}{11169} (\bibinfo{year}{1996}).

\bibitem[{\citenamefont{Bl{\"o}chl}(1994)}]{Blochl:1994uk}
\bibinfo{author}{\bibfnamefont{P.~E.} \bibnamefont{Bl{\"o}chl}},
  \bibinfo{journal}{Phys. Rev. B} \textbf{\bibinfo{volume}{50}},
  \bibinfo{pages}{17953} (\bibinfo{year}{1994}).

\bibitem[{\citenamefont{Iniguez}(2008)}]{Iniguez:2008gm}
\bibinfo{author}{\bibfnamefont{J.}~\bibnamefont{Iniguez}},
  \bibinfo{journal}{Phys. Rev. Lett.} \textbf{\bibinfo{volume}{101}},
  \bibinfo{pages}{117201} (\bibinfo{year}{2008}).

\bibitem[{\citenamefont{Malashevich et~al.}(2012)\citenamefont{Malashevich,
  Coh, Souza, and Vanderbilt}}]{Malashevich:2012et}
\bibinfo{author}{\bibfnamefont{A.}~\bibnamefont{Malashevich}},
  \bibinfo{author}{\bibfnamefont{S.}~\bibnamefont{Coh}},
  \bibinfo{author}{\bibfnamefont{I.}~\bibnamefont{Souza}}, \bibnamefont{and}
  \bibinfo{author}{\bibfnamefont{D.}~\bibnamefont{Vanderbilt}},
  \bibinfo{journal}{Phys. Rev. B} \textbf{\bibinfo{volume}{86}},
  \bibinfo{pages}{094430} (\bibinfo{year}{2012}).

\bibitem[{\citenamefont{Sukhov and Berakdar}(2008)}]{Sukhov:2008cz}
\bibinfo{author}{\bibfnamefont{A.}~\bibnamefont{Sukhov}} \bibnamefont{and}
  \bibinfo{author}{\bibfnamefont{J.}~\bibnamefont{Berakdar}},
  \bibinfo{journal}{J. Phys. Condens. Matter} \textbf{\bibinfo{volume}{20}},
  \bibinfo{pages}{125226} (\bibinfo{year}{2008}).

\bibitem[{\citenamefont{Foner}(1963)}]{Foner:1963vi}
\bibinfo{author}{\bibfnamefont{S.}~\bibnamefont{Foner}},
  \bibinfo{journal}{Phys. Rev.} \textbf{\bibinfo{volume}{130}},
  \bibinfo{pages}{183} (\bibinfo{year}{1963}).

\bibitem[{\citenamefont{Stone and Vickerman}(1971)}]{Stone:1971gj}
\bibinfo{author}{\bibfnamefont{F.~S.} \bibnamefont{Stone}} \bibnamefont{and}
  \bibinfo{author}{\bibfnamefont{J.~C.} \bibnamefont{Vickerman}},
  \bibinfo{journal}{Transactions of the Faraday Society}
  \textbf{\bibinfo{volume}{67}}, \bibinfo{pages}{316} (\bibinfo{year}{1971}).

\bibitem[{\citenamefont{Dzyaloshinskii}(1960)}]{Dzyaloshinskii:1960}
\bibinfo{author}{\bibfnamefont{I.~E.} \bibnamefont{Dzyaloshinskii}},
  \bibinfo{journal}{JETP Lett.} \textbf{\bibinfo{volume}{10}},
  \bibinfo{pages}{628} (\bibinfo{year}{1960}).

\bibitem[{\citenamefont{Astrov}(1960)}]{Astrov:1960vt}
\bibinfo{author}{\bibfnamefont{D.~N.} \bibnamefont{Astrov}},
  \bibinfo{journal}{Sov. Phys. JETP} \textbf{\bibinfo{volume}{11}},
  \bibinfo{pages}{708} (\bibinfo{year}{1960}).

\bibitem[{\citenamefont{Finger and Hazen}(1980)}]{Finger:1980kx}
\bibinfo{author}{\bibfnamefont{L.~W.} \bibnamefont{Finger}} \bibnamefont{and}
  \bibinfo{author}{\bibfnamefont{R.~M.} \bibnamefont{Hazen}},
  \bibinfo{journal}{J. Appl. Phys.} \textbf{\bibinfo{volume}{51}},
  \bibinfo{pages}{5362} (\bibinfo{year}{1980}).

\bibitem[{\citenamefont{Beattie and Gilson}(1970)}]{Beattie:1970ks}
\bibinfo{author}{\bibfnamefont{I.~R.} \bibnamefont{Beattie}} \bibnamefont{and}
  \bibinfo{author}{\bibfnamefont{T.~R.} \bibnamefont{Gilson}},
  \bibinfo{journal}{J. Chem. Soc., A} \textbf{\bibinfo{volume}{5}},
  \bibinfo{pages}{980} (\bibinfo{year}{1970}).

\bibitem[{\citenamefont{Lucovsky et~al.}(1977)\citenamefont{Lucovsky, Sladek,
  and Allen}}]{Lucovsky:1977uu}
\bibinfo{author}{\bibfnamefont{G.}~\bibnamefont{Lucovsky}},
  \bibinfo{author}{\bibfnamefont{R.~J.} \bibnamefont{Sladek}},
  \bibnamefont{and} \bibinfo{author}{\bibfnamefont{J.~W.} \bibnamefont{Allen}},
  \bibinfo{journal}{Phys. Rev. B} \textbf{\bibinfo{volume}{16}},
  \bibinfo{pages}{4716} (\bibinfo{year}{1977}).

\bibitem[{\citenamefont{Shim et~al.}(2004)\citenamefont{Shim, Duffy, Jeanloz,
  Yoo, and Iota}}]{Shim:2004eo}
\bibinfo{author}{\bibfnamefont{S.~H.} \bibnamefont{Shim}},
  \bibinfo{author}{\bibfnamefont{T.~S.} \bibnamefont{Duffy}},
  \bibinfo{author}{\bibfnamefont{R.}~\bibnamefont{Jeanloz}},
  \bibinfo{author}{\bibfnamefont{C.-S.} \bibnamefont{Yoo}}, \bibnamefont{and}
  \bibinfo{author}{\bibfnamefont{V.}~\bibnamefont{Iota}},
  \bibinfo{journal}{Phys. Rev. B} \textbf{\bibinfo{volume}{69}},
  \bibinfo{pages}{4253} (\bibinfo{year}{2004}).

\bibitem[{\citenamefont{Baroni et~al.}(2001)\citenamefont{Baroni, de~Gironcoli,
  Dal~Corso, and Giannozzi}}]{Baroni:2001tn}
\bibinfo{author}{\bibfnamefont{S.}~\bibnamefont{Baroni}},
  \bibinfo{author}{\bibfnamefont{S.}~\bibnamefont{de~Gironcoli}},
  \bibinfo{author}{\bibfnamefont{A.}~\bibnamefont{Dal~Corso}},
  \bibnamefont{and}
  \bibinfo{author}{\bibfnamefont{P.}~\bibnamefont{Giannozzi}},
  \bibinfo{journal}{Rev. Mod. Phys.} \textbf{\bibinfo{volume}{73}},
  \bibinfo{pages}{515} (\bibinfo{year}{2001}).

\bibitem[{\citenamefont{Xiang et~al.}(2011)\citenamefont{Xiang, Kan, Wei,
  Whangbo, and Gong}}]{Xiang:2011cn}
\bibinfo{author}{\bibfnamefont{H.~J.} \bibnamefont{Xiang}},
  \bibinfo{author}{\bibfnamefont{E.~J.} \bibnamefont{Kan}},
  \bibinfo{author}{\bibfnamefont{S.-H.} \bibnamefont{Wei}},
  \bibinfo{author}{\bibfnamefont{M.~H.} \bibnamefont{Whangbo}},
  \bibnamefont{and} \bibinfo{author}{\bibfnamefont{X.~G.} \bibnamefont{Gong}},
  \bibinfo{journal}{Phys. Rev. B} \textbf{\bibinfo{volume}{84}},
  \bibinfo{pages}{224429} (\bibinfo{year}{2011}).

\bibitem[{\citenamefont{Fedorova et~al.}(2015)\citenamefont{Fedorova, Ederer,
  Spaldin, and Scaramucci}}]{Fedorova:2015fu}
\bibinfo{author}{\bibfnamefont{N.~S.} \bibnamefont{Fedorova}},
  \bibinfo{author}{\bibfnamefont{C.}~\bibnamefont{Ederer}},
  \bibinfo{author}{\bibfnamefont{N.~A.} \bibnamefont{Spaldin}},
  \bibnamefont{and}
  \bibinfo{author}{\bibfnamefont{A.}~\bibnamefont{Scaramucci}},
  \bibinfo{journal}{Phys. Rev. B} \textbf{\bibinfo{volume}{91}},
  \bibinfo{pages}{165122} (\bibinfo{year}{2015}).

\bibitem[{\citenamefont{Dudko et~al.}(1971)\citenamefont{Dudko, Eremenko, and
  Semenenko}}]{Dudko:1971fv}
\bibinfo{author}{\bibfnamefont{K.~L.} \bibnamefont{Dudko}},
  \bibinfo{author}{\bibfnamefont{V.~V.} \bibnamefont{Eremenko}},
  \bibnamefont{and} \bibinfo{author}{\bibfnamefont{L.~M.}
  \bibnamefont{Semenenko}}, \bibinfo{journal}{Phys. Status Solidi B}
  \textbf{\bibinfo{volume}{43}}, \bibinfo{pages}{471} (\bibinfo{year}{1971}).

\bibitem[{\citenamefont{Tobia et~al.}(2010)\citenamefont{Tobia, De~Biasi,
  Granada, Troiani, Zampieri, Winkler, and Zysler}}]{Tobia:2010dl}
\bibinfo{author}{\bibfnamefont{D.}~\bibnamefont{Tobia}},
  \bibinfo{author}{\bibfnamefont{E.}~\bibnamefont{De~Biasi}},
  \bibinfo{author}{\bibfnamefont{M.}~\bibnamefont{Granada}},
  \bibinfo{author}{\bibfnamefont{H.~E.} \bibnamefont{Troiani}},
  \bibinfo{author}{\bibfnamefont{G.}~\bibnamefont{Zampieri}},
  \bibinfo{author}{\bibfnamefont{E.}~\bibnamefont{Winkler}}, \bibnamefont{and}
  \bibinfo{author}{\bibfnamefont{R.~D.} \bibnamefont{Zysler}},
  \bibinfo{journal}{J. Appl. Phys.} \textbf{\bibinfo{volume}{108}},
  \bibinfo{pages}{104303} (\bibinfo{year}{2010}).

\bibitem[{\citenamefont{Lichtenberg and Lieberman}(1992)}]{Lichtenberg:1992vb}
\bibinfo{author}{\bibfnamefont{A.~J.} \bibnamefont{Lichtenberg}}
  \bibnamefont{and} \bibinfo{author}{\bibfnamefont{M.~A.}
  \bibnamefont{Lieberman}}, \emph{\bibinfo{title}{{Regular and Chaotic
  Dynamics, 2nd Ed., Applied Mathematical Sciences}}}
  (\bibinfo{publisher}{Springer}, \bibinfo{year}{1992}).

\bibitem[{\citenamefont{Gilbert and Kelly}(2004)}]{Gilbert:2004}
\bibinfo{author}{\bibfnamefont{T.~L.} \bibnamefont{Gilbert}} \bibnamefont{and}
  \bibinfo{author}{\bibfnamefont{J.~M.} \bibnamefont{Kelly}},
  \bibinfo{journal}{IEEE Trans. Magn.} \textbf{\bibinfo{volume}{40}},
  \bibinfo{pages}{3443} (\bibinfo{year}{2004}).

\bibitem[{\citenamefont{Chubykalo-Fesenko
  et~al.}(2006)\citenamefont{Chubykalo-Fesenko, Nowak, Chantrell, and
  Garanin}}]{ChubykaloFesenko:2006hq}
\bibinfo{author}{\bibfnamefont{O.}~\bibnamefont{Chubykalo-Fesenko}},
  \bibinfo{author}{\bibfnamefont{U.}~\bibnamefont{Nowak}},
  \bibinfo{author}{\bibfnamefont{R.~W.} \bibnamefont{Chantrell}},
  \bibnamefont{and} \bibinfo{author}{\bibfnamefont{D.}~\bibnamefont{Garanin}},
  \bibinfo{journal}{Phys. Rev. B} \textbf{\bibinfo{volume}{74}},
  \bibinfo{pages}{094436} (\bibinfo{year}{2006}).

\bibitem[{\citenamefont{Skubic et~al.}(2008)\citenamefont{Skubic, Hellsvik,
  Nordstrom, and Eriksson}}]{Skubic:2008gs}
\bibinfo{author}{\bibfnamefont{B.}~\bibnamefont{Skubic}},
  \bibinfo{author}{\bibfnamefont{J.}~\bibnamefont{Hellsvik}},
  \bibinfo{author}{\bibfnamefont{L.}~\bibnamefont{Nordstrom}},
  \bibnamefont{and} \bibinfo{author}{\bibfnamefont{O.}~\bibnamefont{Eriksson}},
  \bibinfo{journal}{J. Phys. Condens. Matter} \textbf{\bibinfo{volume}{20}},
  \bibinfo{pages}{315203} (\bibinfo{year}{2008}).

\bibitem[{\citenamefont{Sukhov and Berakdar}(2009)}]{Sukhov:2009gn}
\bibinfo{author}{\bibfnamefont{A.}~\bibnamefont{Sukhov}} \bibnamefont{and}
  \bibinfo{author}{\bibfnamefont{J.}~\bibnamefont{Berakdar}},
  \bibinfo{journal}{Phys. Rev. Lett.} \textbf{\bibinfo{volume}{102}},
  \bibinfo{pages}{057204} (\bibinfo{year}{2009}).

\end{thebibliography}

\newpage
\newpage
\appendix*
\section{Appendix}
\subsection{Detailed expressions for the renormalized frequencies}
The explicit expressions for the mode frequencies renormalized by the anharmonic coupling are 
\begin{widetext}
\begin{eqnarray}
	\noIR  &=& \omega_{\mathrm{IR}} -\frac{g^2 \AR^2}{8\oR^2\oIR}+\frac{g^2\AOM^2}{4\oR^2\oIR}-\frac{g^2 \AIR}{16\oIR[\oR^2-4\oIR^2]}-\frac{g^2\AOM^2}{4\oIR[\oR^2-(\Omega+\oIR)^2]}-\frac{g^2\AOM^2}{4\oIR[\oR^2-(\Omega-\oIR)^2]}\nonumber\\
	& &-\frac{g^2\AR^2}{8\oIR [\oIR^2-(\oIR+\oR)^2]}-\frac{g^2\AR^2}{8\oIR [\oIR^2-(\oIR-\oR)^2]}+\frac{3\gamma_\mathrm{IR}\AIR^2}{8\oIR}+\frac{3\gamma_\mathrm{IR}\AOM^2}{4\oIR} \label{eqn_correctionomegaIR}\;,
\end{eqnarray}
for the IR mode, and
\begin{eqnarray}
 \noR =	\omega_\mathrm{R} -\frac{g^2 \AIR^2}{8\oR^2[\oIR^2-(\oIR+\oR)^2]}-\frac{g^2\AIR^2}{8\oR[\oIR^2-(\oIR-\oR)^2]}-\frac{g^2\AOM^2}{8\oR[\oIR^2-(\Omega-\oR)^2]}+\frac{3\gamma_\mathrm{R}\AR^2}{8\oR}\label{eqn_correctionomegaR}\;,
\end{eqnarray}
\end{widetext} 
for the R mode.

\end{document}